\documentclass[aps,prd,onecolumn,groupedaddress,nofootinbib,amssymb]{revtex4}
\usepackage[T1]{fontenc}
\usepackage[latin1]{inputenc}
\usepackage{graphicx}
\usepackage[english]{babel}
\usepackage{amsmath}
\usepackage{amssymb}
\usepackage{amsfonts}

\date{\today}

\def\ba{\begin{eqnarray}}
\def\ea{\end{eqnarray}}
\def\beq{\begin{equation}}
\def\eeq{\end{equation}}

\renewcommand{\epsilon}{\varepsilon}

\def\beqt{\begin{tabular}}
\def\eet{\end{tabular}}
\def\beqf{\begin{figure}}
\def\eef{\end{figure}}
\def\beqa{\begin{eqnarray}}
\def\eeqa{\end{eqnarray}}

\renewcommand{\epsilon}{\varepsilon}

\renewcommand{\epsilon}{\varepsilon}

\def\be{\begin{equation}}
\def\ee{\end{equation}}
\def\bea{\begin{eqnarray}}
\def\eea{\end{eqnarray}}
\def\beq{\begin{eqnarray}}
\def\eeq{\end{eqnarray}}

\def\e{{\rm e}}

\begin{document}


\title{A bird's eye view of $f(R)$-gravity}



\author{Salvatore Capozziello$^1$}
\email[]{capozziello@na.infn.it}
\author{Mariafelicia De Laurentis$^1$}
\email[]{mariafelicia.delaurentis@na.infn.it}
\author{Valerio Faraoni$^2$}
\email[]{vfaraoni@ubishops.ca}

 \affiliation{$^1$ Dip. di Scienze Fisiche, Universit\`a di
Napoli ``Federico~II'' and INFN Sez. di Napoli\\
Compl. Universitario Monte S. Angelo, Ed.~N, Via Cinthia, I-80126
Napoli, Italy}

\affiliation{$2$ Physics Department, Bishop's University\\
Sherbrooke, Qu\'ebec, Canada J1M~1Z7}


\begin{abstract}
The currently observed accelerated expansion of the Universe
suggests that cosmic flow dynamics is dominated by some unknown
form of dark energy characterized by a large negative pressure.
This picture comes out when such a new ingredient, beside baryonic
and dark matter,  is considered as a source in the r.h.s. of the
field equations. Essentially, it should be some form of
un-clustered, non-zero vacuum energy which, together with
(clustered) dark matter, should drive the global cosmic  dynamics.
Among the proposals to explain the experimental situation,  the
``{\it concordance model}'', addressed as $\Lambda$CDM,  gives a
reliable snapshot of the today observed Universe according to the
CMBR, LSS and SNeIa data,  but presents dramatic shortcomings as
the ``{\it coincidence and  cosmological constant problems}''
which point out its inadequacy  to fully trace back the
cosmological dynamics. On the other hand, alternative theories of
gravity,  extending in some way General Relativity, allow to
pursue a different approach  giving rise to suitable cosmological
models where a late-time accelerated expansion can be achieved in
several  ways. This viewpoint does not require to find out
candidates for dark energy and dark matter at fundamental level
(they have not been detected up to now), it takes into account
only the ``observed'' ingredients (i.e. gravity, radiation and
baryonic matter), but the l.h.s. of the Einstein equations has to
be modified.  Despite of this modification, it could be in
agreement with the spirit of General Relativity since the only
request is that the Hilbert-Einstein action should be generalized
asking for a gravitational interaction  acting, in principle,  in
different ways at different scales. We survey the landscape of
$f(R)$ theories of  gravity in their various formulations, which
have been used to model the  cosmic acceleration as alternatives
to dark energy and dark matter. Besides, we take into account the
problem of gravitational waves in such theories.  We discuss some
successes of $f(R)$-gravity (where $f(R)$ is a generic function of
Ricci scalar $R$), theoretical and experimental challenges that
they face in order to satisfy minimal criteria for viability.
\end{abstract}

\maketitle \textit{Keywords}: alternative theories of gravity;
dark energy; dark matter; gravitational radiation.

\section{Introduction}

Theories of gravity, alternative to Einstein's General Relativity
(GR), have been proposed to cure the problems of the standard
cosmological model  and, above all, because they arise in
quantizations of  gravity. These alternative gravitational
theories constitute at least an attempt to formulate a
semi-classical scheme in which  GR and its most successful
features can be recovered. One of the most fruitful approaches
thus far has been that of {\it Extended Theories of Gravity}
(ETGs), which have become a paradigm in the  study of the
gravitational interaction. ETGs are based on corrections and
extensions of  Einstein's theory. The paradigm consists,
essentially, of  adding higher order curvature invariants and/or
minimally or non-minimally coupled scalar fields to the dynamics;
these corrections emerge from  the effective action of quantum
gravity \cite{odintsov}.

Further motivation to modify GR arises from the problem of fully
implementing  Mach's principle in a theory of gravity, which
leads
one to contemplate a varying gravitational coupling. Mach's
principle states that the local inertial frame is determined by
the average motion of distant astronomical objects \cite{bondi}.
This fact would imply   that
the gravitational coupling here and now is determined by the
distant distribution of matter, and it can be scale-dependent
and related to some scalar field. As a consequence,  the concept
of ``inertia'' and the Equivalence Principle have to be revised.
Brans-Dicke theory \cite{BD} constituted the first consistent
and complete theory alternative to Einstein's GR. Brans-Dicke
theory incorporates  a variable gravitational coupling strength
whose dynamics are  governed by a scalar field non-minimally
coupled to the geometry, which implements  Mach's principle
in the gravitational theory \cite{BD, cimento,sciama}.

Independent motivation for extending gravity comes from the fact
that  every unification scheme of the fundamental interactions,
such as Superstring, Supergravity, or Grand Unified Theories
exhibit effective actions containing non-minimal couplings to the
geometry or higher order terms in the curvature invariants. These
contributions are one-loop or higher loop corrections in the
high-curvature regime approaching  the full, and still unknown,
quantum gravity regime \cite{odintsov}.  Specifically, this scheme
was adopted in the study of quantum field theory on curved
spacetime  and it was found that interactions between quantum
scalar fields and background geometry, or gravitational
self-interactions, yield such corrections to the Einstein-Hilbert
Lagrangian \cite{birrell}. Moreover, it has been realized that
these corrective terms are inescapable in the effective action of
quantum gravity close to the Planck energy \cite{vilkovisky}. Of
course, all these approaches do not constitute a full quantum
gravity theory, but are needed as working schemes toward it.

In summary, higher order terms in the invariants of the
Riemann tensor, such as
$R^{2}$, $R^{\mu\nu} R_{\mu\nu}$,
$R^{\mu\nu\alpha\beta}R_{\mu\nu\alpha\beta}$, $R \,\Box R$, or $R
\,\Box^{k}R$, and non-minimal coupling terms between scalar
fields and geometry such as $\phi^{2}R$, have to be added to the
effective gravitational Lagrangian  when
quantum  corrections are introduced. These terms occur also in
the effective Lagrangian of string or Kaluza-Klein theories
when a mechanism of compactification of extra spatial dimensions
is used \cite{veneziano}.

From a conceptual point of view, there is no {\it a priori}
reason to restrict the gravitational Lagrangian to a
linear function of the Ricci scalar $R$ minimally coupled with
matter \cite{francaviglia}. Furthermore, the idea has been
proposed that there are no exact laws of physics, in the sense
that the effective Lagrangians describing  physical interactions
could be stochastic functions at the microscopic level. This
property would imply that local gauge invariances and the
associated  conservation hold only in the low energy limit and
the fundamental constants of physics can vary \cite{ottewill}.

Besides fundamental physics motivations, all these theories
have been the subject of enormous attention in cosmology due to
the fact that they naturally exhibit an inflationary behaviour
which can overcome the shortcomings of  the GR-based
standard cosmological model. The cosmological scenarios arising
from ETGs  seem  realistic and capable of reproducing
observations of the the cosmic
microwave background (CMB)
 \cite{starobinsky,kerner,la}. It has been
shown that, by means of  conformal transformations, the
higher order and non-minimally coupled terms
can be related  to Einstein gravity with one
or more  scalar fields  minimally coupled to9 gravity
\cite{teyssandier,maeda,wands1,wands,gottloeber}.

Higher order terms always  appear as contributions of even order
in the field equations.  For example, the term $R^{2}$ produces
fourth order equations \cite{ruzmaikin}, $R \ \Box R$ gives sixth
order equations \cite{gottloeber,sixth}, $R \,\Box^{2}R$  eighth
order equations \cite{eight}, and so on. By means of a conformal
transformation, any second order derivative term corresponds to a
scalar field.\footnote{The dynamics of these scalar fields are
governed given by a second order Klein-Gordon-like equation.}.
Fourth-order gravity corresponds to Einstein gravity with one
scalar field, sixth-order gravity to Einstein gravity with  two
scalar fields, {\em etc.} \cite{gottloeber,schmidt1} It is also
possible to show that  $f(R)$ gravity is equivalent not only to a
scalar-tensor theory,  but also to GR plus an ideal fluid
\cite{cno}. This feature becomes interesting if  multiple
inflationary events are desired, because an early inflationary
stage could select very large scale structures (observed as
clusters of galaxies today), while a later inflationary epoch
could select smaller scale structures  (observed as galaxies
today) \cite{sixth}, with  each inflationary era corresponding to
the dynamics of a scalar field. Finally, these extended schemes
could naturally solve the graceful exit problem  bypassing the
shortcomings of known inflationary models \cite{la,aclo}.

In addition to the  revision of standard cosmology at early
epochs with the concept of inflation,  a new approach is
necessary also at late epochs.   ETGs could play a fundamental
role also in this context. In fact,  the increasing bulk of data
accumulated in the past few years have nurtured a new
cosmological model referred  to as the
{\it Concordance Model}. The Hubble diagram of type Ia Supernovae
(hereafter SNeIa) measured by both the Supernova Cosmology
Project \cite{SCP} and
the High-$z$ Team \cite{HZT} up to redshifts $z \sim 1$, has been
the first piece of evidence  that the universe is
currently undergoing a
phase of accelerated expansion. Balloon-born
experiments, such as BOOMERanG \cite{Boomerang} and MAXIMA
\cite{Maxima}, have detected  the first and second
peak in the anisotropy spectrum of the CMB
radiation  indicating that the geometry of the
universe is spatially flat. In conjunction with
constraints  on the matter density
parameter $\Omega_M$ coming from galaxy clusters,
these data indicate that the universe is dominated by an
unclustered fluid with negative pressure, generically dubbed
{\it dark energy}, which is able to drive the accelerated
expansion. This picture has been further strengthened by the
recent precise measurements of the CMB spectrum obtained by the
WMAP
experiment \cite{WMAP,hinshaw,hinshaw1}, and by the extension of
the SNeIa Hubble diagram to redshifts higher than one
\cite{Riess04}.
An  overwhelming flood of papers has appeared following
this  observational evidence,
presenting  a great variety of models trying
to explain this phenomenon. The simplest explanation
is the well known cosmological constant $\Lambda$
\cite{LCDMrev}. Although it is the best fit to most of the
available astrophysical data \cite{WMAP}, the $\Lambda$CDM model
fails in explaining why the inferred value of $\Lambda$ is so tiny
(120 orders of magnitude smaller) in comparison with
the typical
vacuum energy values predicted by particle physics and why its
energy density is comparable to the matter density today  (the
{\it coincidence problem}).

As a tentative solution, many authors have replaced the
cosmological constant with a scalar field rolling down its
potential and giving rise to the model referred to as {\it
quintessence} \cite{QuintRev,tsu1}. Even when  successful in
fitting
the data, the quintessence approach to dark energy is still
plagued by the coincidence problem since the dark energy and
matter densities evolve differently and reach comparable values
for a very limited portion of the cosmic  evolution  coinciding
at the present era. To be more precise, the quintessence dark
energy
is tracking matter and evolves in the same way for a long time.
But then, at late times, somehow it has to change its behavior
from tracking the dark matter to  dominating as a
cosmological constant. This is the coincidence problem of
quintessence.

Moreover, the origin of this quintessence scalar field is unknown,
leaving a great uncertainty on the choice of the scalar field
potential. The subtle and elusive nature of  dark energy has led
many authors to  look for completely different scenarios able to
give a quintessential behavior without the need for exotic
components. To this end, it is worth stressing that the
acceleration of the universe only calls for a dominant component
with negative pressure,  but does not tell us anything about the
nature
and the number of cosmic fluids filling the universe. This
consideration suggests that it could be possible to explain
the accelerated expansion by introducing a single cosmic fluid
with an equation of state causing it to act like dark matter at
high densities and dark energy at low densities. An attractive
feature of these models, usually referred to as {\it Unified Dark
Energy} (UDE) or {\it Unified Dark Matter} (UDM) models, is that
such an approach naturally solves, al least phenomenologically,
the coincidence problem. Interesting examples are the
generalized Chaplygin gas \cite{Chaplygin}, the tachyon field
\cite{tachyon} and the condensate cosmology \cite{Bassett}. A
different class of UDE models has been proposed \cite{Hobbit}
in which a single fluid is considered: its energy density scales
with
the redshift in such a way that the radiation-dominated era, the
matter  era, and the accelerating phase can be naturally
achieved. These models are very
versatile since they can be interpreted both in the framework of
UDE models and as a two-fluid scenario with dark matter and scalar
field dark energy. The main advantage of this approach is that a
suitable generalized equation of state can be always obtained and
observational data can be fitted.

There is a yet different way to address the problem of the
cosmic acceleration. As stressed in \cite{LSS03}, it is possible
that the observed acceleration is not the manifestation of another
ingredient of the cosmic pie, but rather the first signal of a
breakdown of our understanding of the laws of gravitation in the
infrared limit. From this point of view, it is tempting to
modify the
Friedmann equations to see whether it is possible to fit the
astrophysical data with  models comprising only standard
matter. Interesting examples of this kind are the Cardassian
expansion \cite{Cardassian} and DGP gravity \cite{DGP}. In the
same framework it is possible to  find alternative
schemes in which a quintessential behavior is obtained by taking
into
account effective models coming from fundamental physics
and giving
rise to generalized or higher order gravity actions
\cite{curvature} (see \cite{odinoj} for a comprehensive review).
For instance, a cosmological constant term may be recovered as a
consequence of a non-vanishing torsion field, leading to a
model consistent with both the SNeIa Hubble diagram and
Sunyaev-Zel'dovich data of galaxy clusters
\cite{torsion}. SNeIa data could also be efficiently fitted
including higher order curvature invariants in the gravitational
Lagrangian \cite{curvfit,camfr}. These alternative models provide
naturally a
cosmological component with negative pressure whose origin is
related to the cosmic geometry, thus overcoming
the
problems linked to the physical significance of the scalar field.

The large  number of cosmological models which constitute viable
candidates to  explain  the observed accelerated expansion is
evident from this short  overview. On the one hand, this
overabundance of models signals the fact that only a
limited number of cosmological tests are available to
discriminate between competing
theories and, on the other hand, it shows that we are facing  an
urgent  degeneracy problem. It is useful to remark that both the
SNeIa Hubble
diagram and the angular size-redshift relation of compact
radio sources \cite{AngTest} are distance-based  probes of
cosmological models, so  systematic errors and biases could be
iterated. From this point of view, it is interesting to search for
tests based on time-dependent observables. For example, one can
take into account the {\it lookback time} to
distant objects since this quantity can discriminate between
different cosmological models. The lookback time is
observationally estimated as the difference between the present
age of the universe and the age of a given object at redshift
$z$. Such an estimate is possible if the object is a galaxy
observed in more than one photometric band since its color is
determined by its age as a consequence of stellar evolution. It is
thus possible to get an estimate of the galaxy age by measuring
its magnitude in different bands and then using stellar
evolutionary codes to choose the model that best reproduces the
observed colors.

Coming to the weak-field-limit approximation, which essentially
means considering Solar System scales,  ETGs are expected to
reproduce GR which, in any case, is firmly tested only in this
limit \cite{Will}. This fact is a matter of debate since several
relativistic theories do not reproduce exactly the Einsteinian
results in the Newtonian approximation but, in some sense,
generalize them. As first noticed by Stelle
\cite{stelle}, an $R^2$-theory gives rise to Yukawa-like
corrections in the Newtonian potential. This feature could have
interesting physical consequences; for example, certain authors
claim
to explain the flat rotation curves of galaxies by using such
terms \cite{sanders}. Others \cite{mannheim} have shown that a
conformal theory of gravity is nothing but a fourth-order
theory containing such terms in the Newtonian limit. Besides,
an apparent, anomalous, long-range acceleration in  the data
analysis of the Pioneer 10/11, Galileo, and
Ulysses spacecrafts could be framed in a general theoretical
scheme by taking  into
account  corrections to the Newtonian potential  \cite{anderson}.

In general, any relativistic theory of gravitation  yields
corrections to the Newtonian and post-Newtonian (PPN) potentials
({\em e.g.}, \cite{schmidt}) which test the theory  \cite{Will}.
Furthermore, the
newborn {\it gravitational lensing astronomy} \cite{ehlers} is
generating  additional tests of gravity over small, large, and
very large scales which  soon will provide direct measurements for
the variation of the Newtonian coupling \cite{krauss}, the
potential
of galaxies, clusters of galaxies and several other features of
self-gravitating systems. Such data, very likely, will be
capable of confirming or ruling
out the physical consistency of GR or of any ETG. In summary, the
general features of ETGs are that the Einstein field equations
are modified in two ways: $i)$ the geometry can be
non-minimally coupled to some scalar field, and/or $ii)$ higher
than second order derivatives of the metric appear. In the
first case we deal with scalar-tensor theories of
gravity; in the second case we have higher order theories.
Combinations of non-minimally coupled and higher order
terms can emerge as contributions to effective Lagrangians;
then we have higher order-scalar-tensor theories of
gravity.

From the mathematical point of view, the problem of reducing
generalized theories to an Einstein-like form has been
extensively discussed. Under suitable regularity
conditions on the Lagrangian and using  a Legendre
transformation on the metric, higher order theories take the
form of GR in which one or more  scalar field(s)
source of the gravitational field (see, {\e.g.},
\cite{francaviglia,sokolowski,ordsup,magnano-soko}). On the other
hand, as discussed above, the mathematical
equivalence between models with variable gravitational coupling
and Einstein  gravity has been studied using suitable
conformal transformations \cite{dicke,nmc}. A debate on the
physical meaning of these conformal
transformations seems to be ongoing (\cite{faraoni} and
references therein). Several authors
claim a physical difference between Jordan frame
(higher order theories and/or variable gravitational couplings)
since there is experimental and observational evidence
suggesting  that the Jordan frame is better suited for
matching solutions and data. Others state that the true physical
frame is the Einstein one according to the energy theorems
\cite{magnano-soko}. However, the discussion is open and no
definitive conclusion seems to have been reached. The problem
becomes more involved at the semiclassical and quantum level, and
should be faced from a more general point of view---the
Palatini approach to gravity could be useful to this goal.

The
Palatini approach to gravitational theories was first
introduced and analyzed by Einstein himself
\cite{palaeinstein}, but was named as a
consequence of an historical misunderstanding
\cite{buchdahl,frafe}.

The fundamental idea  of the Palatini formalism is to consider the
torsion-free connection $\Gamma^{\mu}_{\alpha\beta}$ entering
the definition of the Ricci tensor, to  be independent of the
spacetime metric $g_{\mu\nu}$. The Palatini
formulation of the standard Einstein-Hilbert  theory turns out
to be equivalent
to the purely metric theory. This property   follows from the
fact that the field equations for the connection
$\Gamma^{\mu}_{\alpha\beta}$, considered
to be independent of the metric, produce the Levi-Civita
connection of the metric $g_{\mu\nu}$. As a consequence, there is
no reason to impose the Palatini variational principle instead of the metric
variational
principle in
the Einstein Hilbert theory. However, the situation
changes if we consider the ETGs, which
depend on functions of the curvature invariants (such as $f(R)$
theories)
or couple non-minimally to some scalar field. In these
cases the
Palatini and the metric variational principles provide different
field equations and the theories thus derived differ
\cite{magnano-soko,FFV}. The relevance of the Palatini approach
for cosmological applications in
this framework has been recently demonstrated
\cite{curvature,odinoj,palatinifR}.

From the physical point of view, considering the metric
$g_{\mu\nu}$ and the connection $\Gamma^{\mu}_{\nu\alpha}$
as independent fields means to decouple the metric structure of
spacetime and its geodesic structure (the
connection $\Gamma^{\mu}_{\alpha\beta}$, in general, is  not the
Levi-Civita
connection of $g_{\mu\nu}$). The
causal structure of spacetime is governed by $g_{\mu\nu}$ while
the spacetime trajectories of particles
are governed
by $\Gamma^{\mu}_{\alpha\beta}$. This decoupling enriches the
geometric structure of spacetime and
generalizes the purely metric formalism. This metric-affine
structure of spacetime  is naturally translated, by means of the
Palatini field equations, into a bi-metric structure of
spacetime. Besides the physical metric $g_{\mu\nu}$, another
metric $\tilde{g}_{\mu\nu}$ appears. This new metric is
related, in the case of
$f(R)$ gravity, to the connection. The
connection $\Gamma^{\mu}_{\alpha\beta}$ turns out to be the
Levi-Civita connection of $ \tilde{g}_{\mu\nu}$ and provides
the geodesic structure of spacetime.

For non-minimally coupled interactions in
the gravitational Lagrangian in scalar-tensor theories, the new
metric $ \tilde{g}_{\mu\nu}$ is  related to the non-minimal
coupling; $\tilde{g}_{\mu\nu} $ can be related to a
different geometric and physical aspect of the gravitational
theory. Thanks to the Palatini formalism, the non-minimal
coupling and the scalar field, entering
the evolution of the gravitational fields, are separated from the
metric structure of spacetime. The situation mixes when we
consider the case of higher order-scalar-tensor theories. Due to
these features, the Palatini approach could contribute to
clarify the physical meaning of conformal transformations
\cite{ACCF}.

\vspace{3.mm}

In this review paper,  without claiming for completeness, we want
to give a survey on the formal and phenomenological  aspects of
ETGs in metric and Palatini approaches, considering the
cosmological and astrophysical applications of some ETG models.
The layout is the following. The field equations for generic ETGs
are derived in Sec.\ref{due}. Specifically, we discuss  metric,
Palatini and metric-affine approaches. In Sec. \ref{tre} the e
quivalence of metric and Palatini $f(R)$ gravities with
Brans-Dicke theories are discussed. In Sec. \ref{quattro} we
introduce theoretical and experimental viability of
$f(R)$-gravity. Briefly, we discuss on the correct cosmological
dynamics and on the instabilities for a particular case of $f(R)$.
After we discuss  the precence of ghost fields and the wealk field
limit for metric approach. Finally we consider the growth of
cosmological perturbations and the Chauchy problem. Cosmological
applications are considered in Sec.\ref{cinque}-\ref{sei}. We show
that dark energy and the dark matter can be addressed as
"curvature effects", if ETGs (in particular $f(R)$ theories) are
considered. We work out some cosmological models  comparing the
solutions with data coming from
 observational surveys. As further result in Sec. \ref{sette}. , we show that also the
stochastic cosmological background of gravitational waves can be
"tuned" by ETGs. This fact could open new perspective also in the
problems of detection of gravitational waves which should be
investigated not only in the standard GR-framework. Discussion and conclusions
are drawn in Sec.\ref{otto}.

\section{The three versions of $f(R)$ gravity}
\label{due}

In this survey we focus on $f(R)$ gravity (see
\cite{review} for a more comprehensive discussion and a list of
references, and \cite{otherreviews} for short introductions to
the
subject). In these theories
the Einstein-Hilbert action\footnote{Here $R$ is the Ricci
curvature
of the metric tensor $g_{\mu\nu} $, which has
determinant $g$, $G$ is Newton's
constant, and  $\kappa \equiv 8\pi G$. We mostly follow the
notations of Ref.~\cite{Wald}.}
\begin{equation}
S_{EH}=\frac{1}{2\kappa}\int d^4x \, \sqrt{-g} \,
R+S^{(m)} \end{equation}
is modified to
\be \label{actionmetric}
S=\frac{1}{2\kappa} \int d^4x \, \sqrt{-g} \, f(R)+S^{(m)} \;,
\ee
where $f(R)$ is a non-linear function of its argument
 and $S^{(m)}$ is the matter part of the action.
 Actually, there  are two variational
principles that one can apply to the  Einstein-Hilbert action
in
order to derive Einstein's equations:  the standard metric
variation and a less standard variation  dubbed Palatini
variation. In the latter the metric and the connection are
assumed to be independent variables and one varies the action
with respect to both of them, under  the important
assumption that the matter action does not depend  on the
connection. The choice of the variational principle  is usually
referred to as a formalism, so one can use the terms  metric
(or second order) formalism and Palatini (or first order)
formalism. However, even though  both
variational principles lead to the same field equation  for an
action whose Lagrangian is linear in $R$, this is no  longer
true for a more general action. Therefore, it is  intuitive that
there will be two version of $f(R)$-gravity, according  to which
variational principle or formalism is used.
Indeed this is the case: $f(R)$-gravity in the metric formalism
is called {\em metric $f(R)$-gravity} and $f(R)$-gravity in the
Palatini formalism is called {\em Palatini $f(R)$-gravity}.

Finally, there is actually even a third version of $f(R)$-gravity:
 {\em metric-affine $f(R)$-gravity}. This comes about if one
uses the Palatini variation but abandons the assumption that
the matter action is independent of the connection. Clearly,
metric affine $f(R)$-gravity is the most general of these
theories and reduces to metric or Palatini $f(R)$-gravity if
further assumptions are made. In this section we will present
the actions and field equations of all three versions of $f(R)$-
gravity and point out their difference. We will also clarify
the physical meaning behind the assumptions that discriminate
them.

Then brefly has we show above three versions of $f(R)$-gravity have been studied:\\
\begin{itemize}
\item metric (or second order) formalism;\\
\item Palatini (or first order) formalism;
\end{itemize}
 and
 \begin{itemize}
\item metric-affine gravity.\\
\end{itemize}

These families of theories are discussed in the following.

\subsection{Metric $f(R)$ gravity}

In the metric formalism  the action is
\be  \label{metricaction2}
S_{metric}=\frac{1}{2\kappa}\int d^4 x \, \sqrt{-g} \,
f(R)+S^{(m)} ,
\ee
and its variation with respect to  $g^{\mu\nu}$ yields, after some
manipulations and modulo surface terms,
the field equation
\be \label{metricfieldeqs}
f'(R)R_{\mu\nu}-\frac{f(R)}{2} \,
g_{\mu\nu}=\nabla_\mu\nabla_\nu f'(R) -g_{\mu\nu} \Box
f'(R) +\kappa\, T_{\mu\nu} \;,
\ee
with a prime denoting differentiation with respect to $R$, $\nabla_\mu$ is the covariant derivative associated with the
 Levi-Civita connection of the metric, and $\Box\equiv
 \nabla^\mu\nabla_\mu$. Fourth
order  derivatives of the metric appear  in the first two terms
on the right hand side, justifying the alternative  name ``fourth
order gravity'' used for this class of theories.

By taking the trace of eq.~(\ref{metricfieldeqs}) one obtains
\be \label{tracemetric}
3\Box f'(R)+Rf'(R)-2f(R)=\kappa \, T \;,
\ee
where $T\equiv {T^{\alpha}}_{\alpha}$ is the trace of the
energy-momentum  tensor of matter. This  second order
differential equation for $f'(R)$ is qualitatively different from
the trace of the Einstein equation $R=-\kappa \, T$ which,
instead, constitutes an algebraic relation between $T$ and
the  Ricci scalar, displaying the fact that $f'(R)$ is a
dynamical  (scalar) degree of freedom of the theory.
This is already an indication that the field equations of $f(R)$
theories will admit a larger variety of solutions than
Einstein's theory. As an
example, we mention here that the Jebsen-Birkhoff's theorem,
stating that the
Schwarzschild solution is the unique spherically symmetric vacuum
solution, no longer holds in metric $f(R)$ gravity.
Without going into  details, let us stress that $T=0$ no longer
implies that $R=0$, or is even constant.
Eq.~(\ref{tracemetric}) will turn out to be very useful in
studying various aspects of $f(R)$ gravity, notably its
stability and  weak-field limit. For the moment, let us use it
to make some remarks about maximally symmetric solutions. Recall
that maximally
symmetric solutions lead to a constant Ricci scalar. For $R={\rm
constant}$ and $T_{\mu\nu}=0$, eq.~(\ref{tracemetric}) reduces to
 \be
\label{metftr}
f'(R)R-2f(R)=0,
\ee
 which, for a given $f$, is an algebraic equation in $R$. If $R=0$ is
a root of this equation and one takes this root, then eq.~(\ref{metricfieldeqs})
reduces to $R_{\mu\nu}=0$ and the maximally symmetric solution is
Minkowski spacetime. On the other hand, if the root of
eq.~(\ref{metftr}) is $R=C$, where $C$ is a constant, then
eq.~(\ref{metricfieldeqs}) reduces to $ R_{\mu\nu}= g_{\mu\nu} C/4 $ and
the
maximally symmetric solution is de  Sitter or anti-de Sitter
space
depending
on the sign of $C$, just as in GR with a cosmological
constant.
Another issue that should be stressed is that of energy
conservation. In metric $f(R)$ gravity the matter is  minimally
coupled to the metric. One can, therefore, use the usual arguments based on the invariance of the action under diffeomorphisms of the spacetime manifold
[coordinate transformations  $x^{\mu}\rightarrow
x'^{\mu}=x^{\mu}+
\xi^{\mu} $ followed by a pullback,
with the field $\xi^{\mu} $ vanishing on the boundary of the
spacetime region considered, leave the physics unchanged, see
\cite{Wald} to show that $T_{\mu\nu}$ is
divergence-free. The
same can be done at the level of the field equations: a ``brute
force'' calculation reveals that the left hand side of
 eq.~(\ref{metricfieldeqs}) is divergence-free (generalized Bianchi
identity)  implying that $\nabla_\mu T^{\mu\nu}=0$.

The field equation~(\ref{metricfieldeqs}) can be rewritten
as  form of  Einstein
equations with an effective stress-energy tensor to the right hand side.  Specifically,  as

\be
G_{\mu\nu}=\kappa \left( T_{\mu\nu}+T_{\mu\nu}^{(eff)} \right)
\ee
where
\be\label{effectivetensor}
T_{\mu\nu}^{(eff)}=\frac{1}{\kappa} \left[ \frac{ f(R)-Rf'(R)}{2}\,
g_{\mu\nu}+\nabla_\mu \nabla_\nu f'(R)-g_{\mu\nu} \Box f'(R) \right]
\ee
is an effective energy-momentum tensor constructed with geometric
terms. Since $T_{\mu\nu}^{(eff)}$ is only a formal
energy-momentum tensor, it is not expected to satisfy any
of the energy
conditions deemed reasonable for physical matter, in particular
the  effective energy density cannot be
expected to be positive-definite. An effective
gravitational coupling $G_{eff}\equiv G/f'(R)$  can be defined in
a way analogous to  scalar-tensor gravity. It is apparent that
$f'(R)$ must be positive for the  graviton to carry
positive kinetic energy.

Motivated by the recent cosmological observations, we adopt the
spatially flat  Friedmann-Lemaitre-Robertson-Walker (FLRW)
metric to describe the universe,
\be
ds^2=-dt^2 +a^2(t) \left( dx^2+dy^2+dz^2 \right) \;,
\ee
where $a$ is the scale factor.
Then, the field equations of metric $f(R)$ cosmology become
\begin{eqnarray}
& & H^2=\frac{\kappa}{3f'(R)} \left[
\rho^{(m)}+\frac{Rf'(R)-f(R)}{2}-3H \dot{R} f''(R) \right] \;,\\
&&\nonumber \\
&& 2\dot{H}+3H^2= -\frac{\kappa}{f'(R)} \left[
P^{(m)}+ f'''(R) \left( \dot{R}\right)^2 +2H\dot{R}
f''(R)+\ddot{R}f''(R) \right. \nonumber \\
&& \nonumber \\
&& \left. +\frac{ f(R)-Rf'(R) }{2} \right] \;,
\end{eqnarray}
where $H \equiv  \dot{a}/a$ is the Hubble
parameter and an overdot denotes differentiation with respect to
the comoving
time $t$. The corresponding phase space is a 2-dimensional curved
manifold  embedded in a 3-dimensional space and with a rather
complicated structure  \cite{deSouzaFaraoni}.

\subsection{Palatini $f(R)$ gravity}
In the Palatini version of $f(R)$ gravity, both the metric
$g_{\mu\nu}$ and the  connection
$\Gamma^\mu_{\nu\gamma}$ are regarded as independent variables.
In other words,  the connection is not the metric connection of
$g_{\mu\nu}$. While in GR the metric and Palatini variations
produce the same field equations ({\em i.e.}, the Einstein
equations), for non-linear Lagrangians one obtains
two different sets of field equations.\footnote{By imposing
that the metric and  Palatini variations generate the
same field equations, Lovelock gravity is
selected \cite{Exirifard}. GR  is a special case of Lovelock
theory.}

Palatini $f(R)$ gravity was  proposed as an alternative to dark
energy, on the same footing as metric $f(R)$ models. The original
model advanced for this   purpose was based on the specific
form $f(R)=R-\mu^4/R$ \cite{palatinifR}.

The Palatini action is \be\label{actionPalatini}
S_{Palatini}=\frac{1}{2\kappa}\int d^4 x \, \sqrt{-g} \, f(
\tilde{R}) +S^{(m)}\left[ g_{\mu\nu}, \psi^{(m)} \right] \;, \ee
where a distinction needs to be made between  two different Ricci
tensors contained in the theory. $R_{\mu\nu}$ is constructed from
the metric connection of the (unique) physical metric
$g_{\mu\nu}$, while $\tilde{R}_{\mu\nu}$  is the Ricci tensor of
the non-metric connection  $\Gamma^\mu_{\nu\gamma}$ and defines
the scalar $\tilde{R}\equiv g^{\mu\nu}\tilde{R}_{\mu\nu}$. The
matter part of the action does not depend explicitly from the
connection $\Gamma^{\mu}_{\alpha\beta}$, but only from the metric
and the matter fields, which we  collectively label as
$\psi^{(m)}$.

By varying the Palatini action~(\ref{actionPalatini}) one
obtains the field equation
\be \label{Palatinifieldeq1}
f'(\tilde{R}) \tilde{R}_{\mu\nu}-\frac{ f(\tilde{R})}{2} \,
g_{\mu\nu}=\kappa \,
T_{\mu\nu} \;,
\ee
in which no second covariant derivative of $f'$ appears, in
contrast with eq.~(\ref{metricfieldeqs}). An
independent variation with respect to the connection yields
\be \label{Palatinifieldeq2}
\tilde{\nabla}_\sigma \left( \sqrt{-g} \, f'(\tilde{R}) g^{\mu\nu}
\right)-\tilde{\nabla}_\sigma \left( \sqrt{-g} \, f'(\tilde{R})
g^{\sigma(\mu}\right) \delta^{\nu)}_\gamma =0 \;,
\ee
where $\tilde{\nabla}_\gamma$ denotes the covariant derivative
associated to the (non-metric) connection
$\Gamma^{\mu}_{\alpha\beta}$.

By tracing eqs.~(\ref{Palatinifieldeq1}) and
(\ref{Palatinifieldeq2}) we obtain
\be \label{Palatinitrace}
f'(\tilde{R}) \tilde{R} -2f(\tilde{R})=\kappa \, T
\ee
and
\be \label{Palatinifieldeq3}
\tilde{\nabla}_\gamma \left( \sqrt{-g} \, f'(\tilde{R})
g^{\mu\nu} \right)=0 \;,
\ee
respectively. Eq.~(\ref{Palatinifieldeq3}) is interpreted as
stating that
$\tilde{\nabla}_\gamma$ is the covariant derivative of the
``new'' metric tensor
\be
\tilde{g}_{\mu\nu}\equiv f'( \tilde{R}) g_{\mu\nu}
\ee
conformally related to $g_{\mu\nu}$.  Eq.~(\ref{Palatinitrace})
is an algebraic
(or trascendental, according to the  functional form of $f(R)$)
equation for $f'(\tilde{R})$, not a differential equation
describing its evolution.  Therefore, $f'(R)$  is a non-dynamical
quantity, in contrast to what happens in  metric $f(R)$ gravity.
The lack of dynamics has consequences which are discussed below.
It is possible to eliminate the non-metric connection
from the field equations by rewriting them  as
\begin{eqnarray}
&& G_{\mu\nu}=\frac{\kappa}{f'}\, T_{\mu\nu}-\frac{1}{2}\left(
R-\frac{f}{f'}\right) g_{\mu\nu} +\frac{1}{f'}
\left(\nabla_\mu\nabla_\nu
-g_{\mu\nu}\Box \right) f'\nonumber\\
&&\nonumber \\
&& -\frac{3}{2(f')^2}  \left[
\nabla_\mu f' \nabla_\nu
f' -\frac{1}{2} g_{\mu\nu} \nabla_\gamma f' \nabla^\gamma f'
\right] \;.
\label{Palatinireformulated}
\end{eqnarray}

\subsection{Metric-affine $f(R)$ gravity}

The third family of $f(R)$ theories, metric-affine $f(R) $
gravity \cite{metricaffine}, is characterized by the fact
that also  the  matter part of the action depends explicitly on
the  connection $\Gamma$, as described by the action
\be
S_{affine}=\frac{1}{2\kappa}\int d^4 x \, \sqrt{-g} \, f\left(
\tilde{R} \right) +S^{(m)}\left[ g_{\mu\nu},
\Gamma^\mu_{\nu\gamma}, \psi^{(m)} \right]
\;.
\ee
$\Gamma^{\mu}_{\alpha\beta}$  is possibly a non-symmetric
connection, which would
lead  to torsion associated with matter and to a reincarnation
of torsion theories. The latter were  introduced in view of
elementary particles, rather than cosmology, by coupling the spin
of  elementary particles to the torsion.
The study of metric-affine  $f(R)$ gravity has not  been
completed yet, in particular  its cosmological
consequences have not been fully elucidated. It is for this
reason that our discussion will be limited to  metric
and Palatini $f(R)$ gravity in what follows.

\section{Equivalence of metric and Palatini $f(R)$ gravities with
Brans-Dicke theories}
\label{tre}
In the same way that one can make variable redefinitions in
classical mechanics in order to bring an equation describing a
system to  a more attractive, or easy to handle, form (and in a
 very similar way to changing coordinate systems), one can also
 perform field redefinitions in a field theory, in order to
rewrite the action or the field equations.

There is no unique prescription for redefining the fields of a
theory. One can introduce auxiliary fields, perform
renormalizations or conformal transformations, or even  simply
redefine fields to one's convenience.
It is important to mention that, at least within a classical
perspective such as the one followed here, two theories are
considered to be dynamically equivalent if, under a suitable
redefinition of the
gravitational and matter fields, one can make their field equations
coincide. The  same statement can be made at the level of the
action. Dynamically equivalent theories give exactly the same
results when
describing a dynamical system which falls within the purview of
these theories. There are clear advantages in exploring the
dynamical
equivalence between theories:  we can use results already
derived for one theory in the study of another, equivalent,
theory.

The term `{\it `dynamical equivalence}'' can be considered misleading
in classical gravity. Within a classical perspective, a theory
is fully
described by a set of field equations. When we are referring to
gravitation theories, these equations  describe the
dynamics of gravitating systems. Therefore, two dynamically
equivalent theories
are actually just different representations of the same
theory (which also makes it clear  that all allowed
representations can be used on an equal footing).

The issue of distinguishing between truly different theories and
different representations of the same theory (or dynamically
equivalent theories) is an intricate one. It has  serious
implications
and has been the cause of many misconceptions in the past, especially
when conformal transformations are used in order to redefine the
fields ({\em e.g.,~}the Jordan and Einstein frames in
scalar-tensor
theory).
In what follows, we review the equivalence between metric and
Palatini $f(R)$ gravity with specific theories within the
Brans-Dicke class with a potential.

Metric $f(R)$ gravity is equivalent to an $\omega=0$ Brans-Dicke
theory\footnote{The
Brans-Dicke action for general values of the Brans-Dicke
parameter $\omega$ is
$ S_{BD} = \frac{1}{2\kappa} \int d^4x \, \sqrt{-g} \left[
\phi R -\frac{\omega}{\phi} \,  \nabla^\gamma\phi
\nabla_{\gamma}\phi -V(\phi)
\right] +S^{(m)} $.} when $f''(R) \neq 0$
\cite{BD}, while Palatini modified gravity
is equivalent to one with  $\omega=-3/2$.  The equivalence
has been rediscovered several times over the years, often in the
context of particular theories \cite{STequivalence}.

\subsection{Metric formalism}

It has been noticed quite early that metric quadratic gravity
can be cast into the form of a Brans-Dicke theory  and it did
not take long for these results to be extended to more
general actions which are functions of the Ricci scalar of the
metric .  Let us
present this equivalence in some detail.

We will work at the level of the action but the same approach
can be used
to work directly at the level of the field equations. We begin with
metric $f(R)$ gravity.
Let $f''(R) $ be non-vanishing and consider the
action~(\ref{actionmetric}); by using  the
auxiliary scalar field $\phi =R$, it is easy to see that  the
action
\be \label{equivalentmetric}
S=\frac{1}{2\kappa} \int d^4 x \, \sqrt{-g} \left[ \psi( \phi)R
-V(\phi)
\right] +S^{(m)}
\ee
with
\be
\psi(\phi) = f'(\phi) \;, \;\;\;\;\;\;
V(\phi)=\phi f'(\phi)-f(\phi)
\ee
is equivalent to the previous one. It is trivial that
(\ref{equivalentmetric}) reduces to~(\ref{actionmetric}) if
$\phi=R$. Vice-versa, the variation of~(\ref{equivalentmetric})
with respect to $g^{\mu\nu}$ yields
\be
G_{\mu\nu}=\frac{1}{\psi}\left( \nabla_\mu\nabla_\nu \psi-
g_{\mu\nu}\Box\psi -\frac{V}{2}\, g_{\mu\nu}
\right)+\frac{\kappa}{\psi} \, T_{\mu\nu} \;.
\ee
The variation with respect to $\phi$, instead, gives us
\be
R\, \frac{d\psi}{d\phi} -\frac{dV}{d\phi}=\left( R-\phi
\right)f''(\phi)=0 \;,
\ee
from which it follows that  $\phi=R$ because $f''\neq 0$. The
scalar field  $\phi=R$ is clearly a dynamical quantity which
obeys  the trace equation
\be
3f''(\phi)\Box \phi+3f'''(\phi)\nabla^{\alpha}\phi\nabla_{\alpha}
\phi +\phi
f'(\phi)  -2f(\phi) =\kappa \, T
\ee
and is massive. Its mass squared
\be
m_{\phi}^2=\frac{1}{3} \left( \frac{f_0'}{f_0 ''}-R_0 \right)
\ee
is computed in the analysis of small  perturbations of de Sitter
space (here a zero subscript denotes  quantities evaluated at
the constant curvature $R_0$ of the de Sitter background).
It is convenient to consider, instead of $\phi$, the scalar
$ \psi \equiv f'(\phi)$ obeying the evolution equation
\be
3\Box \psi +2 U(\psi) -\psi\, \frac{dU}{d\psi}=\kappa \, T \;,
\ee
where $ U(\psi)=V(\phi(\psi))-f(\phi(\psi)) $.

To summarize, metric $f(R)$ gravity contains a scalar degree
of freedom and the action
\be
S=\frac{1}{2\kappa} \int d^4 x \, \sqrt{-g} \left[ \psi R -U(\psi)
\right] +S^{(m)} \;,
\ee
is identified as an $\omega=0$ Brans-Dicke theory. This theory
(``massive dilaton gravity'') was  introduced in the
1970's in order to generate a Yukawa term in the Newtonian limit
\cite{OHanlon72}, and then abandoned. The assumption
$f''\neq 0$ is interpreted  as the requirement of
invertibility of the  change of  variable $R\rightarrow \psi(R)$.

\subsection{Palatini formalism}
In Palatini modified gravity the equivalence with
a Brans-Dicke theory is discovered in a way similar  to
that of the metric formalism. Beginning with the
action~(\ref{actionPalatini}) and defining  $\phi
\equiv \tilde{R}$
and $\psi \equiv f'(\phi)$, it is seen that, apart from an
irrelevant
boundary term, the action
can be rewritten as
\be \label{equivalentPalatini}
S_{Palatini}=\frac{1}{2\kappa}\int d^4x \, \sqrt{-g} \left[ \psi R
+\frac{3}{2\psi} \, \nabla^\gamma \psi\nabla_\gamma \psi -V(\psi) \right]
+S^{(m)}
\ee
in terms of the metric $g_{\mu\nu}$ and its
Ricci tensor $R_{\mu\nu}$. Here we have used  the property that,
since  $\tilde{g}_{\mu\nu}=\psi \, g_{\mu\nu}$, the
Ricci curvatures of  $g_{\mu\nu}$ and
$\tilde{g}_{\mu\nu}$ satisfy the relation
\be
\tilde{R}=R +\frac{3}{2\psi}
\nabla^\gamma\psi \nabla_{\gamma}\psi-\frac{3}{2} \Box \psi \;.
\ee
The  action~(\ref{equivalentPalatini}) is easily identified as a
Brans-Dicke theory with Brans-Dicke parameter $\omega=-3/2$.

\section{Theoretical and experimental viability of $f(R)$
gravity}
\label{quattro}

In order  to be acceptable, $f(R)$ theories should not only
reproduce the current acceleration of the universe,  but they
must also satisfy the constraints imposed by Solar System and
terrestrial experiments on relativistic gravity, and they must
obey  certain minimal requirements for theoretical viability.
More precisely, these families of theories must:

\begin{itemize}

\item possess the correct cosmological dynamics;
\item be free from instabilities and ghosts;
\item attain the correct Newtonian and post-Newtonian limits;
\item originate cosmological perturbations compatible with the
observations of the CMB and with large
scale structure surveys;
and
\item possess a well-formulated and well-posed initial value problem.
\end{itemize}

If a  single one of these criteria is not met the theory should
be regarded as unviable. In the following we examine how
$f(R)$ gravity performs with regard to these criteria.

\subsection{Correct cosmological dynamics}
According to the tenets of standard cosmology, an acceptable
cosmological model must contain an  early inflationary era (or
possibly another mechanism) solving  the horizon, flatness, and
monopole problems and generating density perturbations, followed
by a radiation- and then a matter-dominated era. The present
accelerated epoch then begins, possibly explained by $f(R)$
gravity.  The future universe  usually consists of  an eternal de
Sitter attractor, or ends in a Big Rip singularity \cite{abdalla}.
Smooth transitions between  different eras are necessary. The exit
from the radiation era, in particular, was believed to be
impossible in many models \cite{Amendolaetal}, but this proved to
be not true. In fact, exit from the radiation or any era can be
obtained as follows. In the approach dubbed ``designer $f(R)$
gravity'' in \cite{otherreviews}, the desired expansion history of
the universe can be obtained by specifying the desired scale
factor $a(t)$ and integrating an ordinary differential equation
for the function $f(R)$ that produces the chosen $a(t)$
\cite{designerf(R)}. In general, the solution to this ODE is not
unique and can assume a form that appears rather contrived in
comparison with simple forms adopted in most popular models.

\subsection{Instabilities}

The choice $f(R)=R-\mu^4/R$ with $\mu\sim H_0\sim
10^{-33}$~eV is again the prototypical example model to discuss
instabilities.   Shortly after it was advanced as an explanation
of the cosmic acceleration, this model was found to suffer
from the pernicious ``Dolgov-Kawasaki'' instability
\cite{DolgovKawasaki}.  This type of instability
was later shown to be common to any metric $f(R)$ theory with
$f''(R)<0$  (\cite{mattmodgrav})
and the extension
to even more general gravitational theories has been discussed
\cite{Zerbini}. Let us  parametrize the deviations
from GR as
\be
f(R)=R+\epsilon \varphi(R)
\ee
with $\epsilon >0$ a small constant with the
dimensions of a mass squared and $\varphi$ dimensionless. The
trace equation for the Ricci scalar $R$ becomes
\be
\Box R+\frac{\varphi '''}{\varphi ''} \, \nabla^\gamma R
\nabla_\gamma R +\left( \frac{\epsilon \varphi ' -1}{3\epsilon
\varphi ''} \right) R =\frac{\kappa \, T}{3\epsilon \varphi
''}+\frac{2\varphi}{3\varphi ''}
\;.
\ee
By expanding around a de Sitter background and
writing the metric {\em locally} as
\be \label{localdS}
g_{\mu\nu}=\eta_{\mu\nu}+h_{\mu\nu} \;,
\ee
and the scalar $R$ as
\be
R=-\kappa\, T +R_1 \;,
\ee
with $R_1$ a perturbation, the first order trace equation
translates into  the
dynamical equation for $R_1$
\be
\ddot{R}_1 -\nabla^2 R_1 -\frac{2\kappa \varphi '''}{\varphi ''}
\, \dot{T}\dot{R}_1+\frac{2\kappa \varphi '''}{\varphi ''} \,
\vec{\nabla}T \cdot \vec{\nabla}R_1 + \frac{1}{3\varphi ''}\left(
\frac{1}{\epsilon}-\varphi ' \right) R_1=\kappa \, \ddot{T}-\kappa
\nabla^2 T -\frac{ \left( \kappa T \varphi^2 +2\varphi \right)}{3\varphi
''} \;.
\ee
The expression containing $\epsilon^{-1}$ dominates the last term
on the left hand  side, giving the effective mass squared of
$R_1$
\be
m^2 \simeq \frac{1}{3\epsilon \varphi ''} \;.
\ee
Therefore, the theory is  stable if $f''(R)>0$ and unstable if
$f''(R)<0$.   Strictly speaking, GR is excluded by the assumption
$f''\neq 0$, but the well-known stability of this case can easily
be included by writing the stability criterion for metric  $f(R)$
gravity as  $f''  \geq 0$.

To go back to the example model of \cite{DolgovKawasaki}
$f(R)=R-\mu^4/R$, this is unstable because
$f''<0$. The small scale $\mu$ determines the  time scale for the
onset of this instability as  $\sim 10^{-26}$~s
\cite{DolgovKawasaki}, making this an
explosive instability.

A physical interpretation of this stability criterion is the
following  \cite{myinterpretation}:  the
effective gravitational coupling is $G_{eff}=G/f'(R)$ and, if
$dG_{eff}/dR=-f''G/(f')^2>0$ (corresponding to $f''<0$), then
$G_{eff}$ increases with $R$ and a large curvature causes gravity
to become stronger and stronger, which in turn
causes a larger $R$, in a positive feedback loop.  If
instead $dG_{eff}/dR<0$, then a negative feedback  stops the
growth of the
gravitational coupling.

What about Palatini $f(R)$ gravity? Since this formalism
contains only second order
field equations and the trace equation $ f'( \tilde{R}) \tilde{R}
-2f(\tilde{R})=\kappa \, T$  is   not a differential equation but
rather a non-dynamical equation, as noted above, there is no
Dolgov-Kawasaki instability \cite{SotiriouPalatiniinstab}.

The discussion of  metric $f(R)$  instabilities
presented above is based on  the
local expansion~(\ref{localdS}) and, therefore, is
limited to short wavelength modes (compared to  the
curvature radius). However, it can be extended to the longest
wavelengths in the case of a de Sitter background
\cite{mydS}. This extension requires a more complicated formalism
because long modes introduce inhomogeneities and are affected by
the notorious gauge-dependence problems of cosmological
perturbations. A covariant and gauge-invariant formalism is
needed here. One proceeds by assuming  that the background space
is de Sitter and by considering the general action
\be
S=\int d^4 x \, \sqrt{-g} \, \left[ \frac{f \left(\phi, R
\right)}{2} -\frac{\omega(\phi)}{2}\, \nabla^\gamma \phi
\nabla_\gamma \phi -V(\phi) \right]
\ee
containing  $f(R)$ and  scalar-tensor gravity as special cases,
and mixtures of them.  The field equations originating from this
action  become, in a FLRW background space,
\begin{eqnarray}
& & H^2= \frac{1}{3f'}
\left(\frac{\omega}{2}\,
\dot{\phi}^2+\frac{Rf'-f}{2}+V-3H\dot{f}\right)
\;, \\
&&\nonumber \\
&& \dot{H}=\frac{-1}{2f'} \left( \omega \dot{\phi}^2
+ \ddot{f'}-H\dot{f'} \right)  \;, \\
&&\nonumber \\
&& \ddot{\phi}+3H\dot{\phi} +\frac{1}{2\omega}\left(
\frac{d\omega}{d\phi}
\dot{\phi}^2 -\frac{\partial f}{\partial \phi} +2 \,
\frac{dV}{d\phi}
\right)=0 \;.
\end{eqnarray}
de Sitter space is  a solution  of the field equations provided
that  the conditions
\be
6H_0^2 f_0'-f_0 +2V_0=0 \;, \;\;\;\;\;\;\;\;
f_0'=2V_0' \;,
\ee
are satisfied.  An analysis of inhomogeneous perturbations of
small amplitude and arbitrary  wavelengths \cite{mydS} using
the covariant and gauge-invariant Bardeen-Ellis-Bruni-Hwang
formalism \cite{Bardeen}  in Hwang's version
\cite{Hwang} for alternative gravitational theories yields the
stability condition in the zero momentum limit
\be \label{stabilitydS}
\frac{ (f_0')^2 -2f_0 f_0''}{f_0' f_0''} \geq 0 \;,
\ee
This is the stability condition of de Sitter space
in metric $f(R)$~gravity with respect to {\em inhomogeneous}
perturbations and  coincides
with the corresponding stability condition with respect to {\em
homogeneous} perturbations \cite{myinterpretation}.

The
equivalence between  metric $f(R)$ gravity and an $\omega=0$
Brans-Dicke theory holds  also at the level of perturbations;
doubts advanced to this  regard  have now
been resolved. The  stability condition
of de Sitter space with respect to inhomogeneous perturbations
in $\omega=0$ Brans-Dicke theory
is  given again by eq.~(\ref{stabilitydS}), while that for
stability with respect
to  homogeneous perturbations is
\be
\frac{ (f_0')^2 -2f_0 f_0''}{f_0'} \geq 0 \;.
\ee
This inequality is again equivalent to~(\ref{stabilitydS})
if stability against {\em local}  perturbations ({\em i.e.},
$f_0''>0$) is also required. Hence, metric $f(R)$ gravity and
$\omega=0$  Brans-Dicke theory are equivalent also
with regard to perturbations.

Beyond the linear approximation, metric $f(R)$ theories have been
shown to be susceptible to  non-linear instability, potentially
threatening the possibility of constructing models of relativistic
stars in strong $f(R)$ gravity. Inside compact objects with
spherical symmetry, a singularity could develop if  $R$ becomes
large \cite{Frolovetc}. Avoiding this singularity requires some
degree of fine-tuning. Various authors have contended that this
problem can be cured by adding, for example, a quadratic term $
\alpha R^2$  to the action as first \cite{abdalla,bamba}. This
problem needs further study, since it could be the biggest
challenge left for metric $f(R)$ theories.

\subsection{Ghost fields}
Ghosts are massive states of negative norm which ruin unitarity
and appear frequently in attempts  to quantize
Einstein's theory. Fortunately,  $f(R)$ gravity theories are free
of  ghosts. More general ETGs of the form $f\left( R,
R_{\mu\nu}R^{\mu\nu},
R_{\mu\nu\gamma\sigma} R^{\mu\nu\gamma\sigma}, ...
\right) $, in general, are plagued by the presence of  ghosts. A
possible exception under certain conditions
studied in \cite{GBghosts} is provided by theories in which
the extra terms are restricted to appear in the Gauss-Bonnet
combination ${\cal
G}=R^2-4R_{\mu\nu}R^{\mu\nu}+
R_{\mu\nu\gamma\sigma}R^{\mu\nu\gamma\sigma}$,  as in $f=f\left(
R, {\cal G}
\right)$. Then, the field equations reduce to second order
equations without ghosts \cite{ Comelli05,
NavarroVanAcoleyen06}.

\subsection{The weak-field limit for metric $f(R)$ gravity}

After errors and omissions in the early treatments of the
weak-field limit of metric and Palatini modified gravity, a
satisfactory discussion of the  particular model $f(R)=R-\mu^4/R$
in the metric formalism appeared  \cite{Chibaetal06}, followed
by the generalization  to arbitrary forms of  the function $f(R)$
\cite{Olmo07, CSE}.

One studies the  PPN parameter
$\gamma$ which is constrained by light deflection experiments in
the Solar System. The goal consists of finding the weak-field
solution of the field equations and, using this solution,
computing the parameter
$\gamma$. A static, spherically symmetric, non-compact body which
constitutes a perturbation of a
background de Sitter universe is considered, as described by  the
line element
\be
ds^2=-\left[ 1+2\Psi(r)-H_0^2r^2 \right] dt^2 +\left[ 1+2
\Phi(r) +H_0^2 r^2 \right] dr^2 +r^2 d\Omega^2
\ee
in Schwarzschild coordinates, with  $d\Omega^2$ being the line
element on the unit 2-sphere. $\Psi$ and $\Phi$ are
post-Newtonian potentials with  small amplitudes, {\em
i.e.}, $\left|\Psi (r) \right| ,
\left|\Phi (r) \right| <<1$,  and small
(non-cosmological)
scales such that $H_0r <<1$ are considered. The
Ricci scalar is expanded  around the
constant curvature of the background de Sitter space as
$R(r)=R_0+R_1$. The PPN parameter $\gamma$ is then given by $
\gamma =-\Phi(r) / \Psi(r) $ \cite{Will}. The analysis relies
upon three assumptions \cite{CSE}:

\begin{enumerate}
\item $f(R)$ is analytical at $R_0$;
\item $mr<<1$, where  $m$ is the effective mass of the scalar
degree of freedom of the theory. In other words,  this scalar
field
is assumed to be light and with  a range larger than the size
of the Solar System (there are no
experimental constraints on scalars
with range $m^{-1} <0.2$~mm).
\item The matter composing the spherical body has negligible
pressure, $P\simeq 0$ and  $ T=T_0+T_1 \simeq -\rho$.
\end{enumerate}

While it is easy to satisfy the first and the last assumptions,
the second one is more tricky, as discussed below. The trace
equation~(\ref{tracemetric}) turns into \be
\nabla^2 R_1 -m^2 R_1 =\frac{-\kappa \rho}{3f_0''} \;,
\ee
regulating the Ricci scalar perturbation,  where
\be \label{cicci}
m^2 =\frac{  (f_0')^2 -2f_0f_0''}{3f_0'  f_0''}
\ee
is the effective mass squared of the scalar, which
reproduces the expression derived in the gauge-invariant
stability analysis of de Sitter space and in propagator
calculations.

If $mr<<1$, the solution of the linearized field equations is
\begin{eqnarray}
&& \Psi(r)=\frac{-\kappa M}{6\pi f_0'}\, \frac{1}{r} \;, \\
&&\nonumber \\
&& \Phi(r) =\frac{\kappa M}{12 \pi f_0'} \, \frac{1}{r} \;,
\end{eqnarray}
and the PPN parameter $\gamma$  is given by
\be
\gamma =\frac{-\Phi(r)}{\Psi(r)}=\frac{1}{2} \;.
\ee
This value manifestly violates  the  experimental bound
\cite{BertottiIessTortora}
\be
\left| \gamma -1 \right| <2.3 \cdot 10^{-5} \;.
\ee
This violation would mark the demise of metric $f(R)$ gravity
were it not for the fact that the second assumption necessary to
perform this calculation is usually not satisfied. In fact, $mr$
fails to be smaller than unity due to
the {\em chameleon effect}. This effect consists of a
dependence of the effective mass  $m$ on the spacetime curvature
or, alternatively, on the matter density of the surroundings. The
scalar degree of freedom can have a short range (for example, $m
> 10^{-3}$~eV, corresponding to a range $\lambda
< 0.2$~mm) at Solar System densities, escaping the experimental
constraints, and have a  long range at cosmological
densities, which allows it to have an effect on the
cosmological dynamics \cite{NavarroVanAcoleyen06, Faulkneretal06}.
While the chameleon effect may seem a form of fine-tuning, one
should bear in mind that  $f(R)$ gravity is complicated and the
effective range does indeed depend on the
environment. The chameleon mechanism is not arranged, but is
built into the theory and is well-known and
accepted in  quintessence models, in which it was originally
discovered  \cite{chameleon}. It has been studied for many forms
of the function $f(R)$ which pass the observational tests. For
example, the model
\be \label{Faulknermodel}
f(R)=R-\left(1-n \right) \mu^2 \left( \frac{R}{\mu^2} \right)^n
\ee
is compatible with the PPN limits if $ \mu \sim
10^{-50}$~eV$\sim10^{-17} H_0$ \cite{Faulkneretal06}. To
understand how this model can work it is sufficient to note that
a correction $\sim R^n$  to the
Einstein-Hilbert
Lagrangian $R$  with $n<1$ will eventually dominate as
$R\rightarrow 0^{+}$. The model~(\ref{Faulknermodel}) agrees with
the experimental data but could be essentially
indistinguishable from a dark energy  model. Discriminating
between dark energy and $f(R)$ models, or between modified
gravity scenarios should be possible on the basis of  the
growth history of cosmological
perturbations.

\subsection{Growth of cosmological perturbations}

Since the spatially homogeneous and isotropic FLRW metric solves
the field equations of many gravitational theories, the
expansion history of the universe by itself cannot
discriminate between various ETGs. However, the  growth of
structures depends on the theory of gravity considered and has
the potential to achieve this goal.
A typical study is that of Ref.~\cite{SongHuSawicki06}; these
authors postulate an expansion history $a(t)$ characteristic  of
the  $\Lambda$CDM model and find that vector
and tensor modes are not affected by $f(R)$ corrections to Einstein
gravity, to lowest order, and can be neglected, whereas
scalar modes do depend on the theory chosen.
In~\cite{SongHuSawicki06}  the
stability condition $f''(R)>0$ discussed above for  scalar
perturbations  is also recovered. It is found
there that $f(R)$ corrections lower the large
angle anisotropies of the cosmic microwave background and produce
correlations between cosmic microwave background and
galaxy surveys which are different from those obtained in   dark
energy models. A rigorous and mathematically self-consistent  approach to the problem of cosmological perturbations in $f(R)$-gravity as been developed using covariant and gauge-invariant quantities  in \cite{sante1,sante2,sante3}).

The study of structure formation in modified gravity is still uncomplete
 and, most of the times, is  carried out within
specific $f(R)$ models. Insufficient attention has been paid
to the fact that some of these models  are already ruled out
because they contradict  the weak-field limit or the stability
conditions. A similar situation is found in  Palatini  models
which, for this reason, will not be discussed
here with regard to their weak-field limit and cosmological
perturbations.

\subsection{The initial value problem}

A physical theory is required to make predictions and,
therefore, it must have a  well-posed Cauchy problem.   GR
satisfies this requirement for ``reasonable'' forms of matter
\cite{Wald}.  The well-posedness of the initial value problem for
vacuum $f(R)$ gravity was briefly  discussed for special metric
models a long time ago
\cite{Noakes}.  Owing to the
equivalence between $f(R)$ gravity and
scalar-tensor gravity when
$f''(R)\neq 0$,  the initial value
problem of $f(R)$ gravity is reduced to the  one
for Brans-Dicke gravity with $\omega=0$ or $ -3/2$.
 The Cauchy  problem  was shown to be  well-posed  for
particular scalar-tensor theories
in \cite{Cocke, Noakes} but  a general analysis has
been completed only relatively recently \cite{Salgado,
Salgado2}. A separate treatment, however, was necessary
for  $\omega=0, -3/2 $ Brans-Dicke theory.

We begin by defining the basic concepts employed: a system of
$3+1$ equations  is said to be {\em well-formulated}
if it can be written as a
system of equations of only first order  in
both temporal and spatial derivatives. Assume that this
system can be  cast in the full  first order form
\begin{equation}
\partial_t \, \vec{u} + M^i \nabla_i \vec{u}=\vec{S}\left(
\vec{u}\right) ,
\end{equation}
where $\vec{u}$ collectively denotes the fundamental variables
$h_{ij}, K_{ij}$, {\em etc.} introduced below, $M^i$ is
called the {\em characteristic matrix} of the system, and
$\vec{S}\left(  \vec{u} \right)$ describes source terms and
contains only the fundamental variables but not their
derivatives. Then, the initial
value formulation is {\em well-posed} if the system of PDEs is {\em
symmetric hyperbolic} ({\em
i.e.}, the matrices $M^i$ are symmetric)  and {\em strongly
hyperbolic} if $ s_iM^i$ has a real set of
eigenvalues and a complete
set of eigenvectors for  any 1-form $s_i$, and obeys some
boundedness conditions \cite{Solin}.

To summarize the results of \cite{TremblayFaraoni}, the Cauchy
problem for metric $f(R)$ gravity is well-formulated   and is
well-posed in vacuo and with ``reasonable'' forms of matter ({\em
i.e.}, perfect fluids, scalar fields, or the Maxwell field). For
Palatini $f(R)$ gravity,  instead, the  Cauchy problem is
well-formulated  \cite{ijggmp} but not well-posed in general, due
to the presence of higher derivatives of  the matter fields in the
field equations and to the fact that it is impossible to eliminate
them \cite{TremblayFaraoni}. However, as it was remarked in
\cite{Cauchycomments}, the Cauchy problem for Palatini is still
well-posed in vacuo and when the trace of the matter
energy-momentum tensor vanishes or it is a constant. On the other
hand, it is possible to show the well-formulation and the
well-position as soon as the source of the field equations is
perfect-fluid matter \cite{Cauchycomments}.

As an alternative, the  Brans-Dicke theory equivalent to
Palatini $f(R)$ gravity can be mapped into its Einstein frame
representation. In this conformal frame the redefined Brans-Dicke
field couples minimally to gravity and non-minimally to matter
\cite{FaraoniTremblay} and the non-dynamical
role of  this scalar is even  more obvious
\cite{FaraoniTremblay}.

The problems with Palatini $f(R)$ gravity manifest themselves
from a completely different angle when one tries to
match static interior and exterior solutions with spherical
symmetry \cite{BarausseSotiriouMiller}.\footnote{Other problems
of Palatini $f(R)$ gravity were   reported and discussed
in~\cite{Kaloper, PalatiniPLB}.}
The field equations are second order PDEs for the metric
components and, since $f$ is a function of $\tilde{R}$, which
in turn is an algebraic function of $T$ due to
eq.~(\ref{Palatinitrace}),  the right hand side of
eq.~(\ref{Palatinireformulated}) contains second derivatives
of $T$. Now,  $T$ contains derivatives of the matter fields up to
first order, hence eq.~(\ref{Palatinireformulated})  contains
derivatives of the matter
fields up to third order. This property is very different from
the familiar situation of GR and  most of its extensions, in
which the field equations contain only
first order derivatives of  the matter fields. A consequence of
this dependence on lower order derivatives of the matter
fields is that, in these theories the metric is generated by an
integral over the
matter sources and discontinuities in the matter fields and
their derivatives are not accompanied by  unphysical
discontinuities  of the metric. In
Palatini $f(R)$ gravity, instead, the algebraic dependence
of  the metric on the matter fields creates unacceptable
discontinuities  in the metric and singularities  in the
curvature, which were discovered in
\cite{BarausseSotiriouMiller}.  Both the failure
of the initial value problem and the presence  of curvature
singularities with  matter fields can be
ascribed to the non-dynamical nature of the scalar degree of
freedom and to the fact that the latter is related algebraically
to $T$. A possible cure consists of modifying the gravitational
sector of the Lagrangian  in such a way that  the order
of the field equations is raised.

\section{Dark energy as curvature}
\label{cinque}
Let us now show, by some straightforward arguments, how
$f(R)$-gravity can be related to the problem of dark energy.
 The
field equations~(\ref{metricfieldeqs}) may be recast in the
Einstein-like form
\begin{equation}\label{5}
G_{\mu \nu} = R_{\mu\nu}-\frac{1}{2}g_{\mu\nu}R =
T^{(eff)}_{\mu\nu}+T_{\mu\nu}/f^\prime(R)
\end{equation}
with $T^{(eff)}$ given by eq.~(\ref{effectivetensor}) and  in
which   matter couples non-minimally to the geometry through
the term $1/f^\prime(R)$.  As noted above, the appearance of
$f^\prime(R)_{;\mu\nu}$ in $T_{\mu\nu}^{(eff)}$ makes
eq.~(\ref{5}) a fourth order equation (unless $f(R) = R$, in
which case the
curvature stress\,-\,energy tensor $T^{(eff)}_{\alpha \beta}$
vanishes identically  and~(\ref{5}) reduces to the
second order Einstein  equation). As is clear from
eq.~(\ref{5}), the curvature stress-energy tensor
$T_{\mu\nu}^{(eff)} $ formally plays the role of a
source in the field equations and  its effect is the same
as that of an effective fluid of
purely geometrical origin. However, one can also consider
the  Palatini approach  \cite{FFV, metricaffine}, in which the
Einstein equations can still be rewritten as effective Einstein
equations containing  a fluid of geometric origin.

In principle, the scheme outlined above provides all the
ingredients needed to tackle the dark side of the universe.
Depending on the scale  considered, the effective  curvature
fluid can  play the role of both dark matter  and dark energy.
From the cosmological point of view, in the standard framework of
a spatially flat homogenous and isotropic universe, the
cosmological dynamics are  determined by the energy budget
through the Friedmann equations.  In particular, the cosmic
acceleration is achieved when the right hand side
of the acceleration equation remains positive. In units
in which  $8 \pi G = c = 1$ this means
\begin{equation}
\frac{\ddot{a}}{a} = - \frac{1}{6} \left ( \rho_{tot} + 3
P_{tot} \right ) \ , \label{eq: fried2}
\end{equation}
where the subscript $tot$ denotes the sum of the curvature fluid
and the matter contributions to the energy density and pressure.
The acceleration condition  $\ddot{a}>0$ for a dust-dominated
model is
\begin{equation}
\rho_{eff} + \rho_M + 3P_{eff} < 0
\ee
or
\be
w_{eff} < - \frac{\rho_{tot}}{3 \rho_{eff}} \;.
\label{eq: condition}
\end{equation}
Then, the effective quantities
\begin{equation}
\rho_{eff} = \frac{8}{f'(R)} \left \{ \frac{1}{2} \left [ f(R)  -
R f'(R) \right ] - 3 H \dot{R} f''(R) \right \} \ , \label{eq:
rhocurv}
\end{equation}
and
\begin{equation}
w_{eff} = -1 + \frac{\ddot{R} f''(R) + \dot{R} \left [ \dot{R}
f'''(R) - H f''(R) \right ]} {\left [ f(R) - R f'(R) \right ]/2 -
3 H \dot{R} f''(R)}  \label{eq: wcurv}
\end{equation}
play a key role in determining the dynamics of the universe.
To gain insight into the dynamics, one can begin by  neglecting
ordinary  matter and studying the power-law form $f(R) = f_0
R^n$ (with $n$ a real number), which represents a straightforward
generalization of Einstein's GR corresponding  the $n=1$ limit.
This choice yields  power-law solutions for the scale
factor $a(t)$ which provide a good fit to the SNeIa
data and are in  good agreement with the estimated age of the
universe in the range $1.366 < n < 1.376$ \cite{curv-ijmpd}. The
same kind of analysis can be carried out  in the presence of
ordinary matter, but in this case, numerical solution of
the field equations is required. Then, it is still possible
to confront the Hubble flow described by such a model with the
Hubble diagram of SNeIa.
\begin{figure}
\centering\resizebox{7.5cm}{!}{\includegraphics{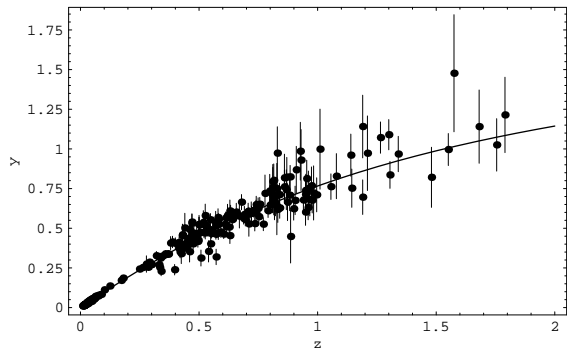}}
\caption{The Hubble diagram of twenty radio galaxies together
with the ``gold" SNeIa sample is plotted versus the redshift $z$,
as  suggested in \cite{daly}. The best-fit curve corresponds to
the  $f(R)$ gravity model without dark matter.  \label{fig:SNeIa}}
\end{figure}
The fit to the data is remarkably good (see
Fig.~\ref{fig:SNeIa}) improving the $\chi^2$ value and it fixes
the best-fit value at $n=3.46$ if the
baryons contribute to the energy density by $\Omega_b \approx
0.04$, in agreement with the prescriptions if Big Bang
nucleosynthesis. The inclusion of dark matter
does not modify the fit appreciably,
supporting the assumption that  dark matter is not essential in
 this model. From the evolution of
the Hubble parameter in terms of redshift, one can even calculate
the age of the universe $t_{univ}$. The best-fit value $n=3.46$
provides
$t_{univ}\approx 12.41$ Gyr. Of course,
$f(R)\,=\,f_0\,R^n$ gravity represents only a toy model
generalization of Einstein's theory. Here we only suggest
that several cosmological and astrophysical results can be well
reproduced  in the realm of a power-law extended gravity model.
This  approach allows flexibility in the value of the exponent
$n$, although it would be preferable to determine a model capable
of working at various scales. Furthermore, we do not expect to
be able to reproduce the entire cosmological phenomenology by
means of a simple power-law model, which is not
sufficiently versatile \cite{Amendolaetal}. For example, it can
be easily demonstrated that this model fails when
it is analyzed with respect to its ability of providing the
correct evolutionary conditions for the perturbation spectra of
matter overdensities \cite{pengjie}. This point is typically
regarded as one of the most important arguments suggesting
the need for darm matter. If one wants to discard this
component, it is crucial to match the observational results
related to the large-scale structure of the universe with the
CMB. These carry the imprints of the initial
matter spectrum at late times
and at early times, respectively.  It is important that the
quantum
spectrum of primordial perturbations, which provide the seeds of
matter perturbations, can be  recovered in the
framework of $R^n$ gravity. In fact, the model $f(R)\,\propto R+
R^2$  can represent a viable model with respect to CMB data and
is a good candidate for cosmological inflation. To obtain the
matter power spectrum suggested by this model, we resort to the
equation for the matter contrast obtained in Ref.~\cite{pengjie}
for fourth order gravity. This equation can be deduced in the
Newtonian conformal gauge for the perturbed metric
\cite{pengjie}
\begin{equation}\label{metric-pert}
ds^2\,=- \left(
1+2\psi \right) dt^2\,+ \,a^2
\left( 1+2\phi \right) \Sigma_{i\,=1}^3(dx^i)^2\,.
\end{equation}
In GR, it is $\phi\,=\,-\psi$ because  there is no anisotropic
stress; in
general,  this relation breaks down in ETGs and
the non-diagonal components of the field equations yield new
relations between the potentials $\phi$ and $\psi$. In  $f(R)$
gravity, due
to the non-vanishing $f_{R;i;j}$ with $i\,\neq\,j$, the
$\phi-\psi$ relation becomes scale-dependent. Instead of the
perturbation equation for the matter contrast $\delta$, we provide
here its evolution in terms of the growth index
$ s\equiv \,d\ln{\delta}/d\ln{a}$,  a quantity
measured at $z\sim 0.15$:
\begin{equation} \label{growind}
s'(a)-\frac{s(a)^2}{a}+
\left[\frac{2}{a}+
\frac{1}{a}E'(a)\right] s(a)
-\frac{1-2Q}{2-3Q}\cdot\frac{3\Omega_m\,a^{-4}}{n\,E(a)^2
\tilde{R}^{n-1}}\,=\,0\,,
\end{equation}
where $E(a)\,=\,H(a)/H_0$, $\tilde{R}$ is the dimensionless
Ricci scalar, and
\begin{equation}\label{Q}
Q\,=\,-\frac{2f_{RR}\,c^2\,k^2}{f_R\,a^2}\,.
\end{equation}
For $n=1$, eq.~(\ref{Q}) gives the
ordinary growth index relation of the Standard Cosmological
Model. It is clear from eq.~(\ref{growind}) that the latter
suggests a dependence of the growth index on the scale which is
contained in the corrective term $Q$ and  that  this dependence
can be safely neglected when
$Q\rightarrow0$. In  the most general case one can
resort to the limit $aH\,<\,k\,<\,10^{-3}h\,Mpc^{-1}$ in which
eq.~(\ref{growind}) is a good approximation, and non-linear
effects on the matter power spectrum can be neglected.

\begin{figure}
{\includegraphics{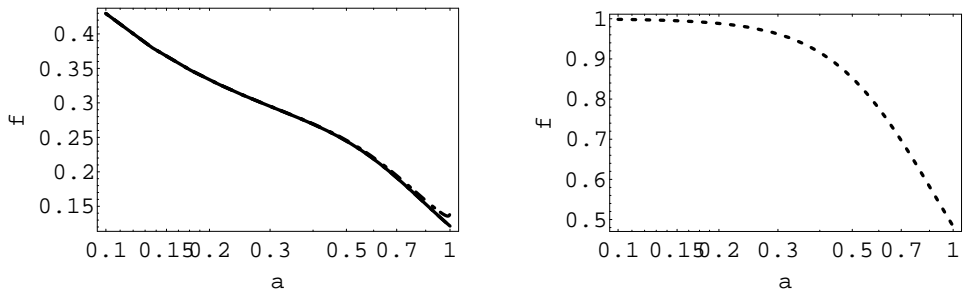}} \caption{The evolution of
the growth index $f$ in terms of the scale factor. The left
panel corresponds to  modified gravity, in the case
$\Omega_m\,=\,\Omega_{bar}\,\sim 0.04$, for the SNeIa best fit
model with $n\,=\,3.46$. The right panel shows the
same evolution in the  $\Lambda$CDM model. In the case
of $R^n$ gravity it is shown also the dependence on the scale
$k$. The three cases
$k\,=\,0.01,\ 0.001$,and $0.0002$  have been eamined, and only
the last of these three cases revelas  a very small deviation
from the leading behavior.
\label{fig: grwf}}
\end{figure}

By studying numerically eq.~(\ref{growind}) one obtains the
 evolution of the growth index in term of the scale factor.
Assuming, for simplicity,  the initial condition
$ s(a_{ls})\,=\,1$ at the last scattering surface as in the case
of matter  domination,  the results are summarized in
Fig.~\ref{fig: grwf},  which displays  the evolution of the
growth  index in $R^n$ gravity and in the $\Lambda$CDM model.

In the case of $\Omega_m\,=\,\Omega_{bar}\,\sim 0.04$, one can
observe a strong disagreement between the expected rate of the
growth index and the behavior induced by power-law fourth order
gravity models. This negative result is evident in  the
predicted value of $ s(a_{z\,=\,0.15})$, which has been
observationally estimated by the analysis of the correlation
function for 220,000 galaxies in the 2dFGRS dataset at the
survey effective depth $z\,=\,0.15$. The observational result
suggests $ s\,=\,0.58\pm0.11$ \cite{lahav}, while our model gives
$ s(a_{z\,=\,0.15})\,\sim\,0.117\ (k\,=\,0.01),\ 0.117\
(k\,=\,0.001),\ 0.122\ (k\,=\,0.0002)$. Although this result seems
frustrating with respect to the underlying idea of discarding the
dark components in the cosmological dynamics, it does not give
substantial improvement in the case of $R^n$ gravity model
plus dark matter. In fact,  it is possible to show that, even in
this case, the growth index prediction is far from being  in
agreement with the $\Lambda$CDM model and again, at the
observational scale $z\,=\,0.15$, there is not enough growth of
perturbations to match the observed large scale structure. In
this case one obtains $ s(a_{z\,=\,0.15})\,\sim\,0.29\
(k\,=\,0.01),\ 0.29\
(k\,=\,0.001),\ 0.31\ (k\,=\,0.0002)$, {\em i.e.}, values which
are  substantially increased with respect to the previous case
but still very far from the experimental estimate. No
significantly different results are obtained if one varies the
power $n$ (of course, for  $n\rightarrow 1$, one recovers
the standard behavior if a cosmological constant is
added to the model). These results seem to suggest that an
extended gravity model incorporating  a simple power-law of the
Ricci scalar, although cosmologically relevant at late times, is
not a viable description of the cosmic evolution  at all
scales. Such a scheme seems too simple to account for the entire
cosmological phenomenology. In \cite{pengjie} a gravity
Lagrangian considering an exponential correction to the Ricci
scalar $f(R)\,=\,R\,+\,A\exp(-B\,R)$ (with $A,\ B$  constants)
produces more interesting results and exhibits a grow factor rate
in agreement with the observational results at least in
the dark matter case. To corroborate this point of view, one has
to consider that when $f(R)$ is chosen  starting
from observational data in an inverse approach as  in
\cite{mimicking}, the reconstructed Lagrangian is a
non-trivial polynomial in the Ricci scalar. This result
suggests that the whole cosmological phenomenology
can be accounted only by a suitable non-trivial function of
the Ricci scalar rather than a simple power-law. The results
obtained in  the study  of the matter power spectra for
simple $R^n$ gravity do not invalidate the general approach.

\section{Dark matter as curvature}
\label{sei}
The results obtained at cosmological scales motivate further
analysis of $f(R)$ theories from the phenomenological point of
view. One  wonders whether the curvature fluid which works as dark
energy could also play the role of effective dark matter,
providing an opportunity  to reproduce the observed astrophysical
phenomenology using only visible matter (see for a discussion
\cite{otherreviews}). It is well known that, in the low energy
limit, higher order gravity implies a modified gravitational
potential, which will play  a fundamental role in our discussion.
By considering a spherical mass distribution with mass $m$ and
solving the vacuum field equations for a Schwarzschild-like
metric, one obtains  the modified gravitational potential of the
theory $f(R)=f_0 R^n $ \cite{mnras}
\begin{equation}
\Phi(r) = - \frac{G m}{2r} \left [ 1 + \left ( \frac{r}{r_c}
\right )^{\beta} \right ]  \;, \label{eq: pointphi}
\end{equation}
where
\begin{equation}
\beta = \frac{ 12n^2 -7n - 1 - \sqrt{ 36n^4 + 12n^3 - 83n^2 + 50n
+ 1 } }{ 6n^2 - 4n + 2 }  \;. \label{eq: bnfinal}
\end{equation}
This potential corrects the ordinary Newtonian potential with  a
power-law term. The correction becomes
important on scales  larger than
$r_c$ and the value of this threshold constant depends
essentially on the mass of the system. The corrected
potential~(\ref{eq: pointphi}) reduces to the
standard Newtonian potential $\Phi \propto 1/r$ for $n=1$, as
follows from the inspection of eq.~(\ref{eq: bnfinal}).

The result~(\ref{eq: pointphi}) deserves some comments. As
discussed in detail in~\cite{mnras}, we have assumed
spherical symmetry in  the the weak-field approximation
of the field equations, which leads to  a corrected Newtonian
potential due to the strong non-linearity of higher order
gravity. Note that Birkhoff's theorem does not hold, in general,
in $f(R)$ gravity,  and that spherically symmetric
solutions different from the Schwarzschild one exist in
these ETGs \cite{noether}.

The generalization of eq.~(\ref{eq: pointphi}) to extended
sources is achieved by dividing the latter into infinitesimal
mass  elements and integrating the potentials generated by
these individual elements. An integral over the
mass density of the system is calculated, taking care of possible
symmetries of the mass distribution \cite{mnras}. Once the
gravitational potential has been computed, one can evaluate the
rotation curve $v_c^2(r)$ and compare it with the
astronomical data. For
extended systems, one typically must resort to numerical
techniques, but the main effect may be illustrated by the rotation
curve for the point-like situation, that is,
\begin{equation}
v_c^2(r) = \frac{G m}{2r} \left [1 + (1 - \beta) \left (
\frac{r}{r_c} \right )^{\beta} \right ] \ . \label{eq: vcpoint}
\end{equation}
In comparison with the Newtonian result $v_c^2 = G
m/r$, the corrected rotation curve is modified by the addition of
the second term on the right hand side of eq.~(\ref{eq:
vcpoint}). For $0 <\, \beta \,< 1$, the
corrected rotation curve is higher than the Newtonian one. Since
measurements of the  rotation curves of spiral galaxies signal
circular velocities larger than predicted by the observed
luminous mass and Newtonian potential,  the above  result
suggests the possibility that the modified gravitational
potential of fourth order gravity may fill the gap between theory
and observations without the need for additional dark matter.

The corrected rotation curve vanishes asymptotically as
in the Newtonian case, while it is usually claimed that observed
rotation curves are flat ({\em i.e.}, asymptotically constant).
Actually, observations do not probe $v_c$ up to infinite radii
but only show  a flat rotation curve (within the
uncertainties) up to the last measured
point. The possibility that $v_c$
goes to zero at infinity is by no means excluded. In order to
check observationally  this result, we have considered a sample
of low surface brightness (LSB) galaxies with well measured HI
and  H$\alpha$ rotation
curves extending far beyond the visible edge of the system. LSB
galaxies are known to be ideal
candidates to test dark matter models because of their
high gas content, which allows the rotation curves to be well
measured and
corrected for possible systematic errors by comparing 21~cm HI
line emission with optical H$\alpha$ and ${\rm [NII]}$ data.
Moreover, these galaxies  are supposed to be
dominated by dark matter, so  fitting their rotation curves
without this elusive component would support ETGs  as
alternatives to dark matter.

Our sample contains fifteen LSB galaxies with data on the
rotation curve, the surface mass density of the gas component and
$R$-band disk photometry extracted from a larger sample
selected by de Blok \& Bosma \cite{BlokBosma}. We assume that
stars
are distributed in an infinitely thin and circularly symmetric
disk with surface density $\Sigma(r) = \Upsilon_\star
I_0$exp${(-r/r_d)}$, where the central surface luminosity $I_0$
and
the disk scale length $r_d$ are obtained from fitting to the
stellar photometry. The gas surface density has been obtained by
interpolating the data over the range probed by HI measurements
and extrapolated outside this range. When fitting the
theoretical rotation curve, there are three quantities to be
determined, namely the stellar mass-to-light ($M/L$) ratio
$\Upsilon_{\star}$, and the theory parameters $\left( \beta,
r_c \right) $. It is
worth stressing that, while fit results for different galaxies
should provide the same value of $\beta$, $r_c$ is related to
one of the integration constants in the field equations. As such,
this quantity  is not  universal and its value must be
determined on a galaxy by galaxy basis. However, it is
expected that galaxies with similar
mass distributions
have similar values of $r_c$ so that the scatter in $r_c$ must
reflect the scatter in the  circular velocities. In order
to match the model with the data, we perform a likelihood analysis
for each galaxy, using as fitting parameters $\beta$,
$\log{r_c}$ (with $r_c$ in kpc) and the gas mass
fraction\footnote{This is related to the $M/L$ ratio
by $\Upsilon_{\star} = [(1 - f_g) M_{g}]/(f_g L_d)$, where w
$M_g = 1.4 M_{HI}$ is the gas (HI + He) mass, and $M_d =
\Upsilon_{\star} L_d$   and
$L_d = 2 \pi I_0 r_d^2$ are the disk total mass and luminosity,
respectively.}
$f_g$. As it is evident from the results of the different
fits, the experimental data are successfully fitted by the model
\cite{mnras}. In particular, from a purely phenomenological
point of view  and leaving aside for the moment other
viability criteria,  from the best fit range $\beta=0.80\pm
0.08$, one can  conclude that $R^n$ gravity with $2.3 < n <5.3$
(best fit value
$n=3.2$ which  overlaps well the above-mentioned range of $n$
fitting SNeIa Hubble diagram) can be a good candidate for solving
the missing matter problem in LSB galaxies without dark matter.
\begin{figure}
\centering\resizebox{7.5cm}{!}{\includegraphics{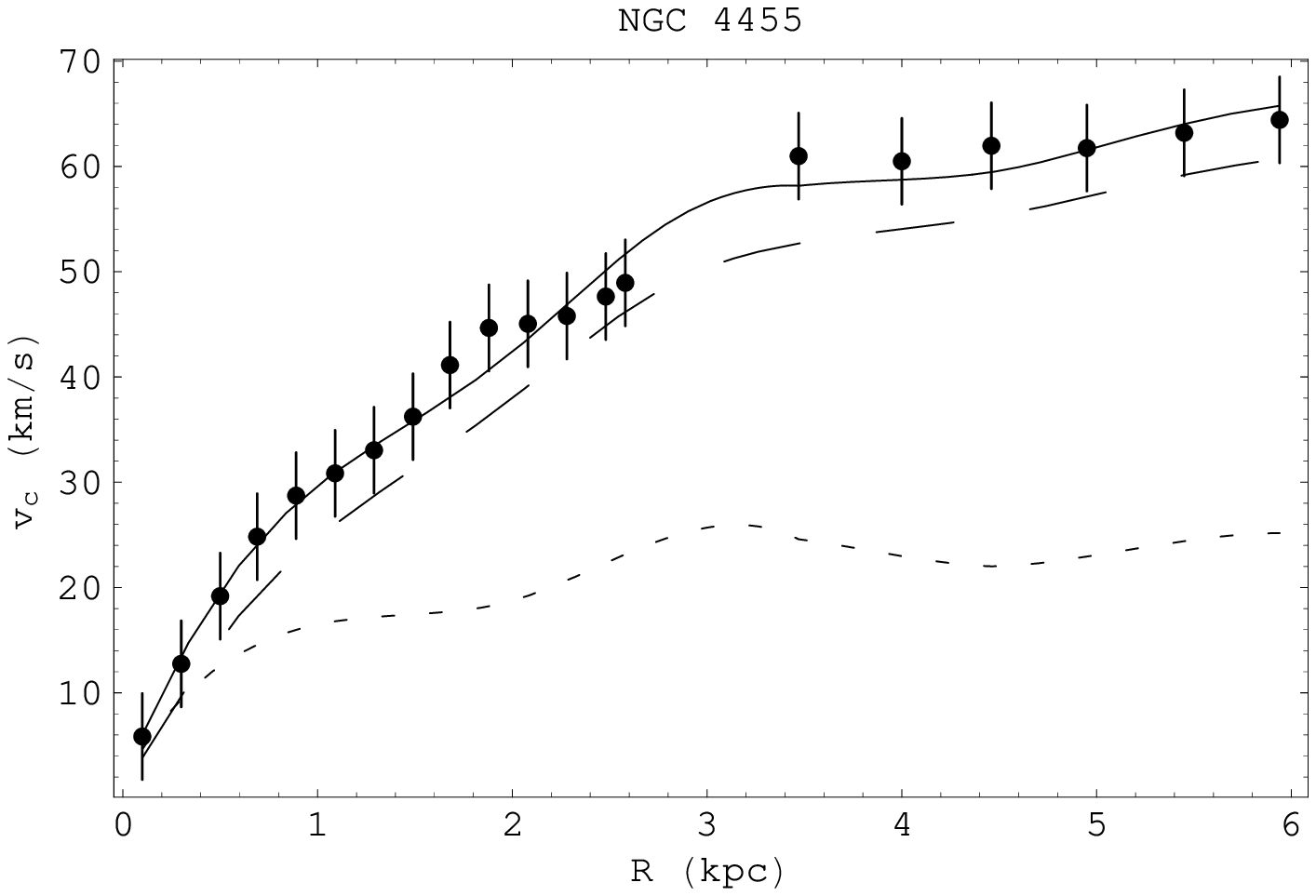}}
\centering\resizebox{7.5cm}{!}{\includegraphics{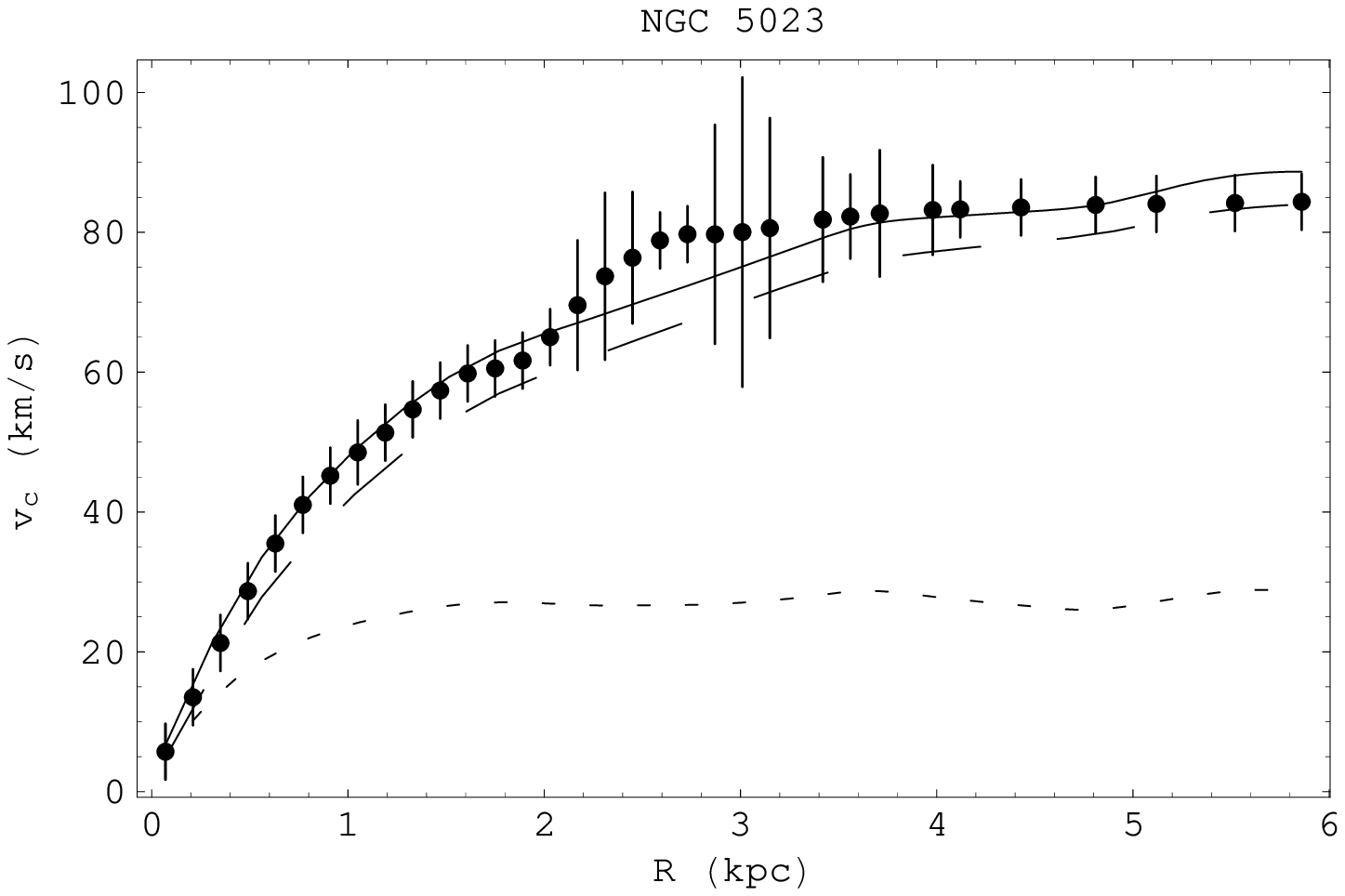}}
\caption{Best-fit theoretical rotation curve superimposed to the
data for the LSB galaxy NGC 4455 (left) and NGC 5023 (right). To
better show the effect of the correction to the Newtonian
gravitational potential, we report the total rotation curve
$v_c(r)$ (solid line), the Newtonian one (short dashed) and the
correction term (long dashed).\label{fig: lsb1}}
\end{figure}

At this point, one wonders whether a link may be found
between $R^n$ gravity and the standard approach based on dark
matter haloes since both theories fit equally well the same data.
As a matter of fact, it is possible to define an {\it effective
dark matter halo} by imposing that its rotation curve equals the
correction term to the Newtonian curve induced by $R^n$ gravity.
Mathematically, one can split the total rotation curve derived
from $R^n$ gravity as $v_c^2(r) = v_{c, N}^2(r) + v_{c,
corr}^2(r)$, where the second term is the correction.
Considering, for simplicity a spherical halo embedding a thin
exponential disk, we may also write the total rotation curve as
$v_c^2(r) = v_{c,
disk}^2(r) + v_{c, DM}^2(r)$ with $v_{c, disk}^2(r)$ the Newtonian
disk rotation curve and $v_{c, DM}^2(r) = G M_{DM}(r)/r$ the dark
matter one, $M_{DM}(r)$ being its mass distribution. Equating the
two expressions yields
\begin{equation}
M_{DM}(\eta) =
M_{vir}\left(\frac{\eta}{\eta_{vir}}
\right)\frac{2^{\beta-5}\eta^{-\beta}_c(1-\beta)
\eta^{\frac{\beta-5}{2}}{\cal I}_0(\eta)-{\cal V}_d(\eta)}
{2^{\beta-5}\eta^{-\beta}_c(1-\beta)\eta^{\frac{\beta-5}{2}}{\cal
I}_0(\eta_{vir})-{\cal V}_d(\eta_{vir})} \label{eq: mdm}
\end{equation}
with $\eta = r/r_d$ and  $\Sigma_0 = \Upsilon_{\star} i_0$,
${\cal
V}_d(\eta)\,=\,I_0(\eta/2)K_0(\eta/2)\times
I_1(\eta/2)K_1(\eta/2)$.\footnote{Here $I_l$ and $K_l$, with
$l\,=\,1,2$ are the Bessel functions of first and second type,
respectively.}
Moreover,
\begin{equation} {\cal{I}}_0(\eta, \beta) =
\int_{0}^{\infty}{{\cal{F}}_0(\eta, \eta', \beta) k^{3 - \beta}
\eta'^{\frac{\beta - 1}{2}} {\rm e}^{- \eta'} d\eta'}
\;, \label{eq:
deficorr}
\end{equation}
where ${\cal{F}}_0$ depends only  on the geometry of the
system and the subscript ``$vir$" indicates virial quantities.
Eq.~(\ref{eq: mdm}) defines the mass profile of an effective
spherically symmetric dark matter halo whose ordinary rotation
curve provides the part of the corrected disk rotation curve
resulting from the addition  of the curvature correction to
the gravitational potential. Clearly, from a phenomenological
point of view there is no way to
distinguish this dark halo model from $R^n$ gravity.

Having assumed spherical symmetry for the mass distribution, it is
straightforward to compute the mass density for the effective
dark halo
as $\rho_{DM}(r) = (1/4 \pi r^2) dM_{DM}/dr$. The most interesting
feature of the density profile is its asymptotic behavior
quantified by the logarithmic slope $\alpha_{DM} =
d\ln{\rho_{DM}}/d\ln{r}$, which can be computed only numerically
as
a function of $\eta$ for fixed values of $\beta$ (or $n$). As
expected, $\alpha_{DM}$ depends explicitly on $\beta$, while
$(r_c, \Sigma_0, r_d)$ enter indirectly through $\eta_{vir}$. The
asymptotic values at the centre and at infinity ($\alpha_0$ and
$\alpha_{\infty}$, respectively) are of  particular
interest. $\alpha_0$ almost vanishes and,  in the
innermost regions, the density is approximately constant. Indeed,
$\alpha_0 = 0$ is  the value corresponding to models with
an isothermal sphere as the inner core. It is well
known that galactic rotation curves are typically best-fitted by
cored dark halo models. Moreover, the outer asymptotic
slope lies between $-3$ and $-2$, values typical of most dark
halo models in the literature. In
particular, for $\beta = 0.80$ one finds $(\alpha_0,
\alpha_{\infty}) = (-0.002, -2.41)$, values which are quite
similar to those in  the Burkert model, $(0, -3)$. This empirical
model  has been proposed to fit  the LSB and dwarf galaxies
rotation curves. The values of $(\alpha_0, \alpha_{\infty})$
found for the best-fit effective dark halo therefore suggest a
possible theoretical motivation for Burkert-like models.
By construction, the properties
of the effective dark matter halo are closely related to the disk
properties, hence  some correlation between the
dark halo
and the disk parameters is expected. In this regard, exploiting
the  relation between the virial mass and the disk parameters,
one obtains the relation for the Newtonian virial velocity
$V_{vir} = G M_{vir}/r_{vir}$
\begin{equation}
M_d \propto \frac{(3/4 \pi \delta_{th} \Omega_m
\rho_{crit})^{\frac{1 - \beta}{4}} r_d^{\frac{1 + \beta}{2}}
\eta_c^{\beta}}{2^{\beta - 6}
 (1 - \beta) G^{\frac{5 - \beta}{4}}} \frac{V_{vir}^{\frac{5 -
\beta}{2}}}{{\cal{I}}_0(V_{vir}, \beta)} \label{eq: btfvir} \ .
\end{equation}
We have checked numerically  that eq.~(\ref{eq: btfvir}) may be
well approximated by $M_d \propto V_{vir}^{a}$. This relation has
the same formal structure of the baryonic Tully-Fisher (BTF)
relation $M_b \propto V_{flat}^a$ where $M_b$ is the total (gas
plus   stars) baryonic mass and $V_{flat}$ is the circular
velocity on  the flat part of the observed rotation curve. In
order to test whether the BTF can be explained by  the
effective dark matter halo proposed, we should look for a
relation between $V_{vir}$ and $V_{flat}$. Such a relation
cannot be derived
analytically because  the estimate of $V_{flat}$ depends on
the peculiarities of the observed rotation curve, such as how
far
it extends, and the uncertainties
on the outermost points. For given values of the disk parameters,
we  simulated theoretical rotation curves for some values
of $r_c$ and measured $V_{flat}$ finally choosing the fiducial
value for $r_c$ that gives a value of $V_{flat}$ as close as
possible to the measured one. Inserting the relation thus found
between $V_{flat}$ and $V_{vir}$ into eq.~(\ref{eq: btfvir}) and
averaging over different simulations, we finally obtain
\begin{equation}
\log{M_b} = (2.88 \pm 0.04) \log{V_{flat}} + (4.14 \pm 0.09) \;,
\label{eq: btfour}
\end{equation}
while a direct fit to the observed data gives \cite{ssm}
\begin{equation}
\log{M_b} = (2.98 \pm 0.29) \log{V_{flat}} + (3.37 \pm 0.13) \ .
\label{eq: btfssm}
\end{equation}
The slope of the predicted and observed BTF are in good
agreement, lending  further support to our approach. The
zero point
is markedly different from the predicted one, being significantly
larger than the observed one. However, both relations fit the
data with a similar scatter. A discrepancy in
the zero point can be due to our approximate treatment of the
effective halo which does not take into account the gas component.
Neglecting this term, we should increase the effective halo mass
and hence $V_{vir}$ which affects the relation with $V_{flat}$
leading to a higher than observed zero point. Indeed, the larger
$M_g/M_d$, the more the points deviate from our predicted BTF thus
confirming our hypothesis. Given this caveat, we can conclude
with some confidence that $R^n$ gravity offers a theoretical
foundation even for the empirically found BTF relation.

Although the results outlined here pertain to the
simplistic choice $f(R)\,=\,f_0R^n$ of fourth order gravity,
they are nevertheless  interesting. The incompatibility
of this  model with the correct matter power spectrum
suggests that a more complicated
Lagrangian is needed to reproduce the entire dark sector
phenomenology at all scales, but it has been  shown that
ETGs allow one to approach  important
issues in  cosmological and astrophysical phenomenology. We have
seen that ETGs can reproduce the SNeIa Hubble
diagram without dark matter and predict  the age of the universe.
The  modification of the gravitational potential
arising in higher order gravity could constitute a
fundamental ingredient in interpreting the flatness of
the rotation curves of LSB galaxies.
Furthermore, if one considers the model parameters selected by
the fit of the observational data of LSB rotation curves, it is
possible to construct a phenomenological analog of the dark
matter halo with shape  similar to that of the Burkert
model. Since the latter has been empirically introduced to
account for  the dark matter distribution in LSB
and dwarf galaxies, this result provides a theoretical
motivation of the Burkert model.

By investigating the relation between dark halo and disk
parameters, a relation has been deduced between $M_d$ and
$V_{flat}$, which reproduces the baryonic Tully-Fisher law.
Exploiting the relation between the virial mass and the disk
parameters, one can obtain a relation for the virial velocity
which can be satisfactorily approximated as $M_d \propto
V_{vir}^{a}$. This result is also  intriguing because  it
provides a theoretical interpretation of another phenomenological
relation. Although not definitive, these
phenomenological aspects of $f(R)$ point to  a potentially
interesting avenue of research and support the quest for
a unified view of the dark side of the universe.

\section{Massive scalar modes of $f(R)$ gravitational waves}
\label{sette}
As we have seen, a  pragmatic point of view could be to
``reconstruct'' the suitable theory of gravity starting from data.
The main issues of this ``inverse '' approach is matching
consistently observations at different scales and taking into
account wide classes of gravitational theories where ``ad hoc''
hypotheses are avoided. In principle, as discussed in the previous
section,  the most popular dark energy cosmological models can be
achieved by considering $f(R)$ gravity without considering unknown
ingredients. The main issue to achieve such a goal is to have at
disposal suitable datasets at every redshift. In particular, this
philosophy can be taken into account also for the cosmological
stochastic background of gravitational waves (GW) which, together
with CMBR, would carry, if detected, a huge amount of information
on the early stages of the Universe evolution.
In this section we discuss the cosmological background of
gravitational waves (GWs) in generic  $f(R)$ theories. The
achievement of detecting massive modes or selecting
$f(R)$-signatures in the stochastic background could be the final
way to retain or rule out such theories with respect to GR.  GWs
are perturbations $h_{\mu\nu}$ of the metric which transform as
3-tensors.  In GR, the equations ruling the propagation of GWs in
the transverse-traceless gauge are
\begin{equation}
\square h_{i}^{j}=0\label{eq: 1}\,,
\end{equation}
where Latin indexes run from~1 to~3. We want to derive
the analog of eq.~(\ref{eq: 1}) for a generic $f(R)$ theory
described by the action~(\ref{actionmetric}). The  linearized
theory in vacuo  ($\mathcal{S}^{(m)}=0$) is considered below, so
that
\begin{equation}
\mathcal{S}=\frac{1}{2k}\int d^{4}x\sqrt{-g}f(R)\label{eq:2}\,.
\end{equation}
Using a  conformal transformation, the scalar degree of
 freedom $f'(R)$ of metric $f(R)$ gravity appears as the
conformal factor in
\begin{equation}
\widetilde{g}_{\mu\nu}=e^{2\Phi}g_{\mu\nu} \, \qquad
\qquad e^{2\Phi}=f'(R)\,.\label{eq:3}
\end{equation}
The conformally equivalent Einstein-Hilbert action is
\begin{equation}
\mathcal{\widetilde{S}}=
\frac{1}{2k}\int d^{4}x \sqrt{-\tilde{g}} \, \left[\widetilde{R}+
\mathcal{L}\left(\Phi\mbox{,}
\Phi_{\mbox{;}\mu}\right)\right] \;, \label{eq:4}
\end{equation}
where $ \mathcal{L} \left(\Phi\mbox{,} \Phi_{\mbox{;}
\mu}\right)$
is the  scalar field Lagrangian obtained using the
relation
\begin{equation}
\widetilde{R}=
e^{-2\Phi}\left(R-6\square\Phi-
6\Phi_{;\delta}\Phi^{;\delta}\right)\label{eq:6}
\end{equation}
between the Ricci curvatures of the conformally related metrics
$g_{\mu\nu}$ and $ \tilde{g}_{\mu\nu}$.
The equation for the gravitational waves is now
\begin{equation}
\widetilde{\square}\tilde{h}_{i}^{j}=0 \;, \label{eq:7}
\end{equation}
where
\begin{equation}
\widetilde{\square}=e^{-2\Phi}\left(\square+
2\Phi^{;\lambda}\partial_{;\lambda}\right)
\,.\label{eq:9}
\end{equation}
Since scalar and tensor modes are decoupled, we have
\begin{equation}
\widetilde{h}_{i}^{j}=\widetilde{g}^{lj}\delta\widetilde{g}_{il}=e^{-2\Phi}g^{lj}e^{2\Phi}\delta
g_{il}=h_{i}^{j} \;, \label{eq:8}
\end{equation}
which means that $h_{i}^{j}$ is conformally invariant. As a
consequence, the plane wave amplitudes
$h_{i}^{j}(t)=h(t)e_{i}^{j}\exp(ik_{m}x^{m}),$ where $e_{i}^{j}$
is the polarization tensor, are the same in both metrics, a fact
that is important in the following.

In a FLRW background,  eq.~(\ref{eq:7}) becomes
\begin{equation}
\ddot{h}+
\left(3H+2\dot{\Phi}\right) \dot{h}+k^{2}a^{-2}h=0\label{eq:10}
\end{equation}
where  $k$ is the wave number and $h$ is the amplitude.
The solutions of this equation are linear combinations of
Bessel functions. Several primordial mechanisms generating GWs
are possible.  In principle, one could seek for contributions
due to all known high-energy  processes in the early phases of
the cosmic history.

Here we consider the  background of GWs generated during
inflation, which is strictly related to the dynamics of the
cosmological model.  In particular, one can assume that the main
contribution to this background comes from the
amplification of vacuum fluctuations at the transition between the
inflationary phase and the radiation  era. However,  we can assume
that the GWs generated as zero-point fluctuations during
inflation undergo adiabatically damped $(\sim 1/a)$ oscillations
until they reach the Hubble radius $H^{-1}$. This is the particle
horizon for the growth of perturbations. Any previous
fluctuation is smoothed away by the inflationary expansion. The
GWs freeze out for $a/k\gg H^{-1}$ and re-enter the horizon
after reheating. The re-entry in the Friedmann era
depends on the spatial scale of the GWs. After re-entry, GWs
are in principle  detectable due the   Sachs-Wolfe effect that
they induce on the CMB temperature anisotropy
$\bigtriangleup T/T$ at decoupling. If $\Phi$ is the
inflaton field, then $\dot{\Phi}\ll H$ during inflation.
By using the conformal time $\eta$ defined by $d\eta=dt/a$,
eq.~(\ref{eq:10}) becomes
\begin{equation}
h''+2\frac{\chi'}{\chi}h'+k^{2}h=0 \;, \label{eq:16}
\end{equation}
where $\chi=ae^{\Phi}$ and  a prime now denotes differentiation
with respect to $\eta$.  Inside the  radius $H^{-1}$, it is
$k\eta\gg 1$.  Since there are no gravitons in the
initial vacuum state, only negative-frequency modes appear and
the solution of eq.~(\ref{eq:16}) is
\begin{equation}
h=k^{1/2}\sqrt{2/\pi}\frac{1}{aH}C\exp(-ik\eta)\,,\label{eq:18}
\end{equation}
where  $C$ is the amplitude parameter. At the first horizon
crossing $aH=k$, the averaged amplitude
$A_{h}=(k/2\pi)^{3/2}\left|h\right|$ of the perturbation is
\begin{equation}
A_{h}=\frac{C}{2\pi^{2}} \,.\label{eq:19}
\end{equation}
When the scale $a/k$ becomes larger than the Hubble radius
$H^{-1}$, the growing  mode  freezes.    It can be shown that
the upper limit  $\bigtriangleup T/T \lesssim  A_{h} $ is valid
since other effects
can contribute to the background anisotropy. From this
consideration, it is clear that the only relevant quantity is the
initial amplitude $C$ in eq.~(\ref{eq:18}), which is conserved
until re-entry. This  amplitude depends  on the fundamental
mechanism that generates the perturbations. Inflation
produces perturbations as zero-point energy
fluctuations, a mechanism which depends on the gravitational
interaction and $(\bigtriangleup T/T)$ further constrains
the theory of gravity.
Considering a single graviton
in the form of a monochromatic wave, its zero-point amplitude is
obtained from  the canonical commutation relation
\begin{equation}
\left[h(t,x),\,\pi_{h}(t,y)\right]=i\delta^{3}(x-y) \label{eq:20}
\end{equation}
at fixed time $t$, where the amplitude $h$ is the field and
$\pi_{h}$ is the conjugate momentum operator. The Lagrangian for
the $h$-quantity is
\begin{equation}
\widetilde{\mathcal{L}}=
\frac{1}{2}\sqrt{-\widetilde{g}} \, \widetilde{g}^{\mu\nu}
h_{;\mu}h{}_{;\nu} \label{eq:21}
\end{equation}
in the conformally rescaled FLRW metric $\widetilde{g}_{\mu\nu}$,
where the amplitude $h$ is conformally invariant. This
Lagrangian leads to
\begin{equation}
\pi_{h}=\frac{\partial\widetilde{\mathcal{L}}}{
\partial\dot{h}}=e^{2\Phi}a^{3}\dot{h}\label{eq:22}
\end{equation}
and eq.~(\ref{eq:20}) becomes
\begin{equation}
\left[h(t,x),\,\dot{h}(y,y)\right]
=i\frac{\delta^{3}(x-y)}{a^{3}e^{2\Phi}} \;.\label{eq:23}
\end{equation}
The fields $h$ and $\dot{h}$ can be expanded in terms of
creation and annihilation operators. The commutation relations in
conformal time are
\begin{equation}
\left[hh'^{*}-h^{*}h'\right]=\frac{i(2\pi)^{3}}{a^{3}e^{2\Phi}}
\,.\label{eq:26}
\end{equation}
Eqs.~(\ref{eq:18}) and~(\ref{eq:19}) yield
$ C=\sqrt{2}\pi^{2}He^{-\Phi}$, where $H$ and $\Phi$ are
calculated at the first horizon crossing and, using
$e^{2\Phi}=f'(R)$, the relation
\begin{equation}
A_{h}=\frac{H}{\sqrt{2f'(R)}}   \label{eq:27}
\end{equation}
is found to hold  for a generic $f(R)$ theory. This result
deserves some discussion. Clearly, the GW amplitude  produced
during inflation depends on the theory of gravity
which, if different from GR, contains extra degrees of freedom
which could be probed by the Sachs-Wolfe effect. This effect
could be combined with other  constraints on the GW  background
if ETGs  are probed independently  at other scales
\cite{tuning, SFM}.

\begin{figure}
\begin{tabular}{|c|c|}
\hline
\includegraphics[scale=0.7]{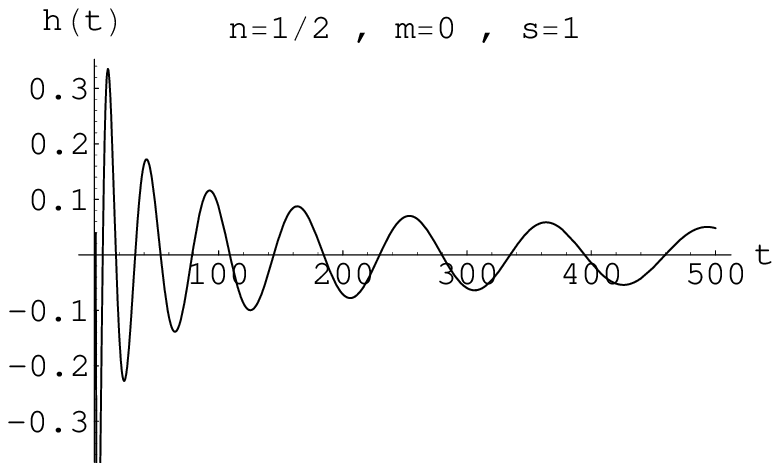}&
\includegraphics[scale=0.7]{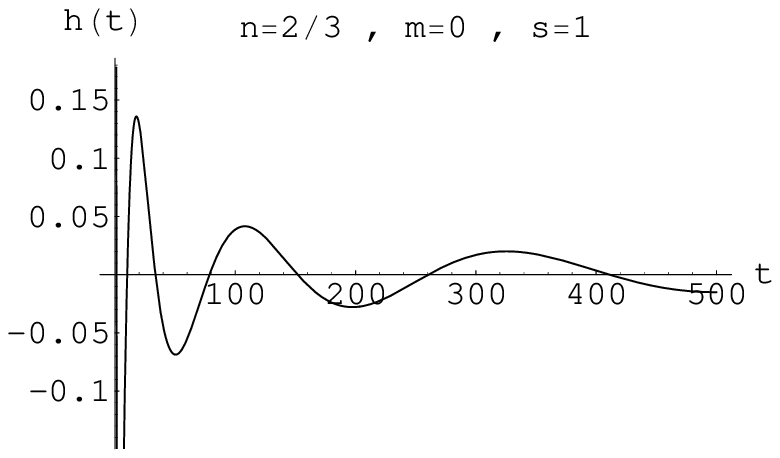}\tabularnewline
\hline
\includegraphics[scale=0.7]{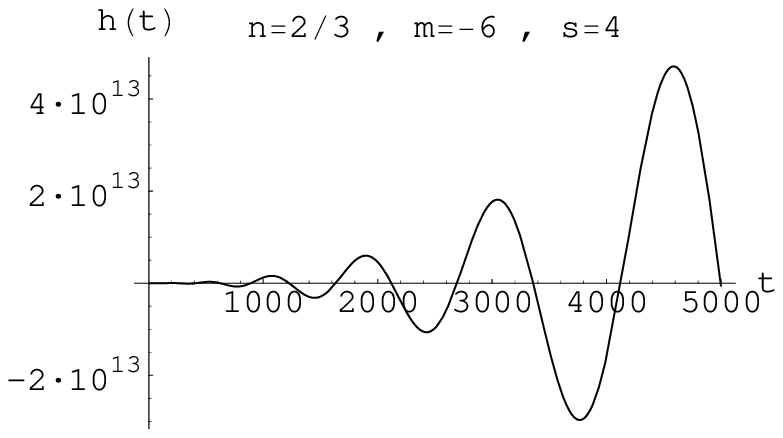}&
\includegraphics[scale=0.7]{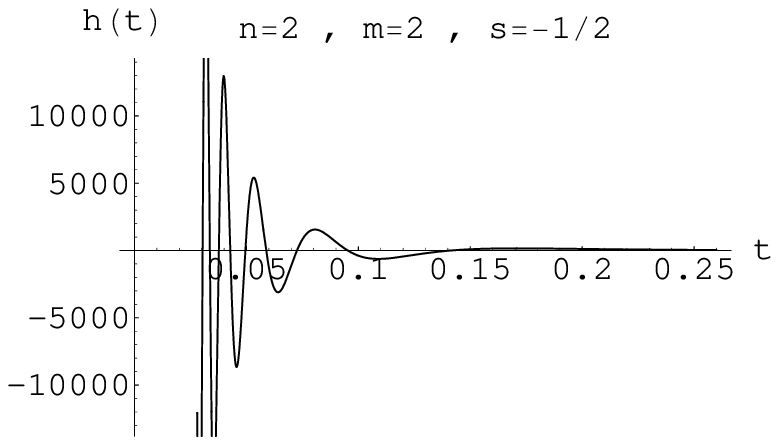}\tabularnewline
\hline
\includegraphics[scale=0.7]{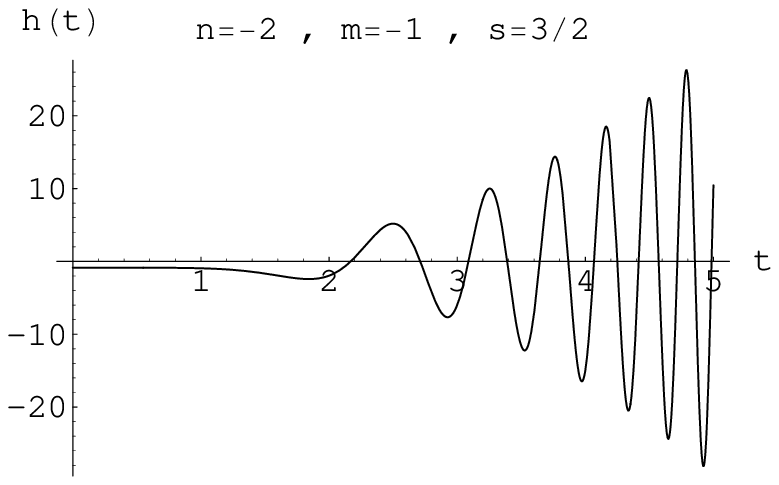}&
\includegraphics[scale=0.7]{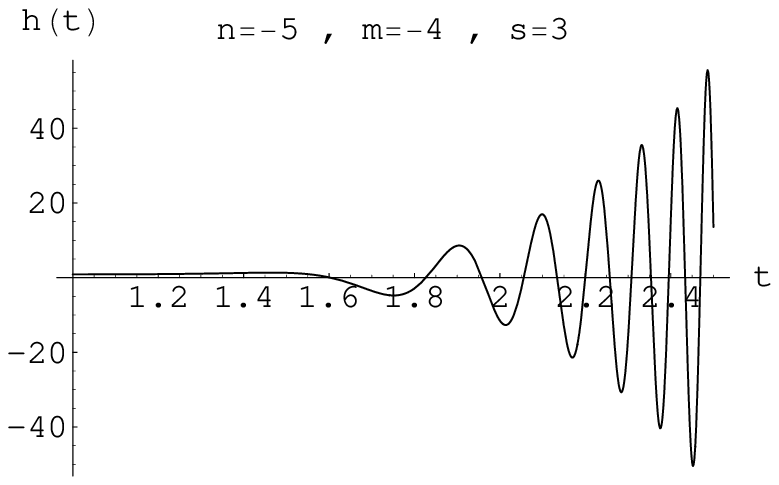}\tabularnewline
\hline
\end{tabular}
\caption {The evolution of the GW amplitude for a few  power-law
choices  of the scale factor $a(t)\sim t^s$, the scalar
field $\phi\sim t^m$, and the function  $f(R)\sim R^n$.
The horizontal (time)  and vertical (amplitude) scales depend on
the cosmological background providing  a signature of the
model.}
\label{fig:1}
\end{figure}

We are by now familiar with the trace of the field  equations
\begin{equation}
3\square f'(R)+Rf'(R)-2f(R)=0,\label{eq: KG}
\end{equation}
and, using the identifications \cite{SCF}
\begin{equation}
\begin{array}{ccccc}
\Phi\rightarrow f'(R) &  &
\textrm{and } &  & \frac{dV}{d\Phi}
\rightarrow\frac{2f(R)-Rf'(R)}{3}
\end{array}\label{eq: identifica}
\end{equation}
the Klein-Gordon equation for the effective  scalar field $\Phi$
\begin{equation}
\square\Phi=\frac{dV}{d\Phi}  \label{eq: KG2}
\end{equation}
follows.  Linearizing around  a constant curvature
background  corresponding to $\Phi=\Phi_{0}$,
assuming that  $V$ has a minimum at $\Phi_{0}$  \cite{SCF},
and expanding as in
\begin{equation}
V\simeq\frac{1}{2}\alpha\delta
\Phi^{2} \;, \;\;\;\;\;\; \frac{dV}{d\Phi}\simeq m^{2}\delta\Phi
];,
\label{eq: minimo}
\end{equation}
where the  constant $m$ has the dimensions of a mass, yields
\begin{equation}
\begin{array}{c}
g_{\mu\nu}=\eta_{\mu\nu}+h_{\mu\nu} \;, \\
\nonumber \\
\Phi=\Phi_{0}+\delta\Phi \;,
\end{array}\label{eq:
linearizza}
\end{equation}
to first order in $h_{\mu\nu}$ and $\delta\Phi$. If
$\widetilde{R}_{\mu\nu\rho\sigma}$,
$\widetilde{R}_{\mu\nu}$, and $\widetilde{R}$ are the linearized
quantities corresponding  to $R_{\mu\nu\rho\sigma}$,
$R_{\mu\nu}$, and $R$, then the linearized field equations are
\begin{equation}
\begin{array}{c}
\widetilde{R}_{\mu\nu}-
\frac{1}{2} \eta_{\mu\nu} \widetilde{R} =
\partial_{\mu}\partial_{\nu}h_{f} -\eta_{\mu\nu}\square h_{f}
\;, \\
\\{}\square h_{f}=m^{2}h_{f} \;,
\end{array}
\label{eq:linearizzate1}
\end{equation}
where
\begin{equation}
h_{f}\equiv\frac{\delta\Phi}{\Phi_{0}} \;.\label{eq:
definizione}
\end{equation}
The curvature tensor $\widetilde{R}_{\mu\nu\rho\sigma}$ and
eqs.~(\ref{eq:linearizzate1})
are left invariant under the gauge transformations
\begin{equation}
\begin{array}{c}
h_{\mu\nu}\rightarrow h'_{\mu\nu}=h_{\mu\nu}
-\partial_{(\mu}\epsilon_{\nu)} \;, \\
\\\delta\Phi\rightarrow\delta\Phi'
=\delta\Phi \;.\end{array}\label{eq: gauge}
\end{equation}
By introducing
\begin{equation}
\bar{h}_{\mu\nu}\equiv
h_{\mu\nu}-\frac{h}{2}\eta_{\mu\nu}+ \eta_{\mu\nu}h_{f}\label{eq:
ridefiniz}
\end{equation}
and considering the gauge vector
$\epsilon^{\mu}$ given by
\begin{equation}
\square\epsilon_{\nu}=\partial^{\mu}
\bar{h}_{\mu\nu} \;,\label{eq:lorentziana}
\end{equation}
the Lorenz gauge
\begin{equation}
\partial^{\mu}\bar{h}_{\mu\nu}=0 \label{eq:
condlorentz}\end{equation}
can be chosen.  In this gauge the field equations assume the form
\begin{equation}
\square\bar{h}_{\mu\nu}=0 \;,  \label{eq: ondaT}
\end{equation}
\begin{equation}
\square h_{f}=m^{2}h_{f} \;.\label{eq: ondaS}\end{equation}
The solutions of eqs.~(\ref{eq: ondaT}) and~(\ref{eq: ondaS}) are
the plane waves
\begin{equation}
\bar{h}_{\mu\nu}=
A_{\mu\nu}(\overrightarrow{p})
\exp(ip^{\alpha}x_{\alpha})+ \mbox{c.c.}  \;, \label{eq:
solT}
\end{equation}
\begin{equation}
h_{f}=a(\overrightarrow{p})\exp(iq^{\alpha}x_{\alpha})
+\mbox{c.c.} \;,
\label{eq: solS}\end{equation}
with
\begin{equation}
\begin{array}{ccc}
k^{\alpha}\equiv(\omega,\overrightarrow{p})
 &  & \omega=p\equiv|\overrightarrow{p}|\\
\\q^{\alpha}\equiv(\omega_{m},\overrightarrow{p})
 &  & \omega_{m}=\sqrt{m^{2}+p^{2}}\;.
\end{array}\label{eq:
keq}\end{equation}
Eqs.~(\ref{eq: ondaT}) and~(\ref{eq: solT}) are the
wave equation  for  standard GR and its
gravitational wave solutions, respectively, whereas
eqs.~(\ref{eq: ondaS}) and~(\ref{eq: solS})
are the wave  equation and its solution for the massive
scalar mode of $f(R)$ gravity (cf. \cite{SCF,BCDF}). The
dispersion relation  for the modes of the
massive
field $h_{f}$ is non-linear. ``Ordinary'' ({\em i.e.},  GR)
tensor modes $\bar{h}_{\mu\nu}$ propagate at the speed of light
$c$, but the dispersion law (the second of eqs.~(\ref{eq: keq}))
for the scalar modes $h_{f}$ is that of a massive field
wave packet \cite{SCF,BCDF}. The group velocity of a
wave packet of $h_{f}$ centered on $\overrightarrow{p}$ is
\begin{equation}
\overrightarrow{v_{G}}=
\frac{\overrightarrow{p}}{\omega}\;,
\label{eq: velocita' di
gruppo}\end{equation}
which is the velocity of a massive particle with mass $m$
and momentum $\overrightarrow{p}$. The second of eqs.~(\ref{eq:
keq}) in conjunction with eq.~(\ref{eq: velocita' di gruppo})
yields
\begin{equation}
v_{G}=\frac{\sqrt{\omega^{2}-m^{2}}}{\omega} \label{eq:
velocita' di gruppo 2}
\end{equation}
and a wave packet propagates at constant speed if
\begin{equation}
m=\sqrt{(1-v_{G}^{2})} \, \omega \;.
\label{eq: relazione
massa-frequenza}
\end{equation}
The  Lorenz gauge is preserved by gauge trasformations with
$\square\epsilon_{\nu}=0$; this gauge imposes the
transversality condition $k^{\mu}A_{\mu\nu}=0$ for the tensor
modes, but not for the field  $h_{\mu\nu}$ which contains a
scalar mode, as seen from  eq.~(\ref{eq: ridefiniz}), or
\begin{equation}
h_{\mu\nu}=\bar{h}_{\mu\nu}-
\frac{\bar{h}}{2}\eta_{\mu\nu}+ \eta_{\mu\nu}h_{f}.\label{eq:
ridefiniz 2}
\end{equation}
Were the scalar mode  massless, one could impose that
\begin{equation}
\begin{array}{c}
\square\epsilon^{\mu}=0 \;, \\
\\\partial_{\mu}\epsilon^{\mu}=
-\frac{\bar{h}}{2}+h_{f} \;,
\end{array}\label{eq: gauge2}\end{equation}
thus obtaining a transversal ``total'' field. However, as is
clear from the previous sections, we are  dealing with a massive
scalar mode and transversality is impossible. By applying
d'Alembert's operator to the second of eqs.~(\ref{eq:
gauge2}) and using eqs.~(\ref{eq: ondaT}) and~(\ref{eq: ondaS}),
it follows that
\begin{equation}
\square\epsilon^{\mu}=m^{2}h_{f} \;,\label{eq:
contrasto}\end{equation}
in contrast with the first of eqs.~(\ref{eq: gauge2}).
Similarly, it is shown that a  linear
relation between the tensorial modes $\bar{h}_{\mu\nu}$  and the
massive scalar $h_{f}$ cannot exist. Thus, a gauge in wich
$h_{\mu\nu}$ is purely spatial cannot be chosen, {\em i.e.},  it
is impossible to impose  $h_{\mu0}=0,$ see eq.~(\ref{eq:
ridefiniz 2}).  However, the traceless gauge condition can
be imposed on $\bar{h}_{\mu\nu}$,
\begin{equation}
\begin{array}{c}
\square\epsilon^{\mu}=0 \;,\\
\\\partial_{\mu}\epsilon^{\mu}=
-\frac{\bar{h}}{2} \;,\end{array}
\label{eq: gauge
traceless}
\end{equation}
implying that
\begin{equation}
\partial^{\mu}\bar{h}_{\mu\nu}=0 \;.\label{eq:
vincolo}\end{equation}
The gauge transformations
\begin{equation}
\begin{array}{c}
\square\epsilon^{\mu}=0 \;, \\
\\\partial_{\mu}\epsilon^{\mu}=0 \;, \end{array}\label{eq: gauge
3}\end{equation}
preserve the gauge $\partial_{\mu}\bar{h}^{\mu\nu}=0$,
$\bar{h}=0$. By choosing $\overrightarrow{p}$ along the
$z$-direction, a gauge can be chosen in which only $A_{11}$,
$A_{22}$, and  $A_{12}=A_{21}$ are different from zero, with
the condition $\bar{h}=0$ providing  $A_{11}=-A_{22}$. The
substitution  of these
equations into eq.~(\ref{eq: ridefiniz 2}) then yields
\begin{equation}
h_{\mu\nu}(t,z)=A^{+}(t-z)e_{\mu\nu}^{(+)}
+A^{\times}(t-z)e_{\mu\nu}^{(\times)}+h_{f}(t-v_{G}z)\eta_{\mu\nu}
\;.\label{eq: perturbazionetotale}\end{equation}
The term $A^{+}(t-z)e_{\mu\nu}^{(+)}+A^{\times}(t-z)e_{\mu\nu}^{
(\times)}$ describes the two standard polarizations of
tensor gravitational waves familiar from GR,  while the term
$h_{f}(t-v_{G}z)\eta_{\mu\nu}$ is the massive scalar field
characteristic of $f(R)$ gravity. As expected, the scalar $f'(R$)
generates a  third massive polarization for gravitational
waves which is absent in GR.

\section{Conclusions}
\label{otto}

Let us emphasize  once more  that we regard $f(R) $ gravity
theories not as definitive theories, but rather as  toy
models and proofs of principle  that modifying gravity at
large scales can explain  the
observed acceleration of the universe without the need to
advocate exotic  dark energy. This hope has stimulated a very
intense activity among theoreticians (\cite{review} and
references therein).

To summarize the status of modified
gravity, let us note that  metric  $f(R)$ gravity  models exist
that pass all the observational and theoretical constraints (see,
{\em e.g.},  the Starobinsky model \cite{SongHuSawicki06} $
f(R)=R+\lambda R_0 \left[ \frac{1}{\left( 1+ \frac{R^2}{R_0^2}
\right)^n }-1 \right] $).
The viable models require the chameleon mechanism in order  to
pass the weak-field limit tests.

All metric $f(R)$ theories must satisfy the  condition $ f''(R)
\geq 0$  to avoid the Dolgov-Kawasaki local instability. This is
a condition on short-wavelength modes.  The stability
condition~(\ref{stabilitydS}) is valid for arbitrary wavelengths,
but is restricted to  de  Sitter space (which is, anyway, an
adiabatic approximation for slowly expanding FLRW spaces). An
important open problem is
whether curvature singularities appear, in general, in
relativistic strong field stars.

As far as the Palatini formalism is concerned,  the
central  idea  of this version of modified gravity  is to
regard the  torsion-free connection $\Gamma^{\mu}_{\alpha\beta}$
as a quantity
independent of the spacetime metric $g_{\mu\nu}$. The
Palatini  formulation of the standard Hilbert-Einstein  theory is
 equivalent to the purely metric theory, as a consequence of the
fact that the field equations for the connection  give the
Levi-Civita connection of the metric $g_{\mu\nu}$. Therefore,
there is  no reason to impose the Palatini variational principle,
instead of the  metric  variational principle,  in the
Einstein-Hilbert theory. However, the situation is difefrent in
ETGs containing non-linear functions of  the curvature invariants,
such as $f(R)$, or non-minimally coupled scalars. In these cases,
the Palatini and the metric variational principle yield different
field equations and different physics \cite{magnano-soko,FFV}. The
relevance of the  Palatini approach for cosmological applications
has been amply  demonstrated
\cite{curvature,odinoj,palatinifR,palatinicam1}. However, Palatini
$f(R)$  theories could have some problems due to the fact that
they could contain  non-dynamical scalar field and the initial
value problem could be ill-posed. In any case, when the trace of
the matter energy-momentum tensor  vanishes identically or it is a
constant, and when it can be recast in a perfect-fluid form, the
Cauchy problem results well-formulated and well-posed.

Metric-affine gravity  has  not been developed in
sufficient detail to assess its viability according to all the
criteria presented  here, and its cosmological consequences are
essentially unexplored.

It seems fair to say that  $f(R)$ theories of gravity can help to
progress in our understanding of the peculiarities of GR in the
wider  landscape of   relativistic theories of gravity.
Furthermore, these theories point out important aspects of
generalizations of GR, and, from a phenomenological point of view,
constitute viable alternatives to dark energy models in explaining
the cosmic acceleration, and to dark matter in  reproducing
dynamical features as the  galactic rotation curves or the halo of
clusters of galaxies \cite{salzano}. Finally, it is possible to
"tune" the stochastic background of GWs and this occurrence could
constitute a further cosmological test capable of confirming or
ruling out ETGs once data from interferometers, like VIRGO, LIGO
and LISA, will be available (see \cite{corda} for a discussion on
this topic). At present, no definite prediction sets $f(R)$
theories apart from dark energy and other models once and for all,
but it is hoped that progress will me made in this direction.

\section*{Acknowledgements}

SC and MD acknowledge V. Cardone, M. Francaviglia, A. Troisi and
S. Vignolo for discussions and some common results used in this
review paper. VF acknowledges financial support from Bishop's
University and the Natural Sciences and Engineering Research
Council of Canada (NSERC).

\end{document}